\documentclass[preprint]{aastex}
\usepackage{bm}

\shorttitle{The effect of an offset polar cap dipolar $B$-field on the modeling of the Vela pulsar's $\gamma$-ray light curves}
\shortauthors{Barnard, Venter, \& Harding}

\begin{document}

\title{The effect of an offset polar cap dipolar magnetic field on the modeling of the Vela pulsar's $\gamma$-ray light curves}

\author{M. Barnard\altaffilmark{1,3} and C. Venter\altaffilmark{1}}
\and
\author{A. K. Harding\altaffilmark{2}}
\altaffiltext{1}{Centre for Space Research, North-West University, Potchefstroom Campus, Private Bag X6001, Potchefstroom 2520, South Africa}
\altaffiltext{2}{Astrophysics Science Division, NASA Goddard Space Flight Center, Greenbelt, MD 20771, USA}
\altaffiltext{3}{Formerly Monica Breed}

\begin{abstract}
We performed geometric pulsar light curve modeling using static, retarded vacuum, and offset polar cap (PC) dipole $B$-fields (the latter is characterized by a parameter $\epsilon$), in conjunction with standard two-pole caustic (TPC) and outer gap (OG) emission geometries. The offset-PC dipole $B$-field mimics deviations from the static dipole (which corresponds to $\epsilon=0$). In addition to constant-emissivity geometric models, we also considered a slot gap (SG) $E$-field associated with the offset-PC dipole $B$-field and found that its inclusion leads to qualitatively different light curves. Solving the particle transport equation shows that the particle energy only becomes large enough to yield significant curvature radiation at large altitudes above the stellar surface, given this relatively low $E$-field. Therefore, particles do not always attain the radiation-reaction limit. Our overall optimal light curve fit is for the retarded vacuum dipole field and OG model, at an inclination angle $\alpha=78{_{-1}^{+1}}^{\circ}$ and observer angle $\zeta=69{_{-1}^{+2}}^{\circ}$. For this $B$-field, the TPC model is statistically disfavored compared to the OG model. For the static dipole field, neither model is significantly preferred. We found that smaller values of $\epsilon$ are favored for the offset-PC dipole field when assuming constant emissivity, and larger $\epsilon$ values favored for variable emissivity, but not significantly so. When multiplying the SG $E$-field by a factor of 100, we found improved light curve fits, with $\alpha$ and $\zeta$ being closer to best fits from independent studies, as well as curvature radiation reaction at lower altitudes.
\end{abstract}


\section{INTRODUCTION}

The field of $\gamma$-ray pulsars has been revolutionized by the launch of the \emph{Fermi} Large Area Telescope (LAT; \citealp{Atwood2009}). Over the past eight years, \textit{Fermi} has detected over 200 $\gamma$-ray pulsars and has furthermore measured their light curves and spectral characteristics in unprecedented detail. \textit{Fermi}'s Second Pulsar Catalog (2PC; \citealp{Abdo2013}) describes the properties of some 117 of these pulsars in the energy range 100~MeV$-$100~GeV. This catalog includes the Vela pulsar \citep{Abdo2009}, one of the brightest persistent sources in the GeV sky. Recently, \textit{H.E.S.S.} detected pulsed emission from the Vela pulsar in the 20$-$120~GeV range (A. Abramowski et al. 2016, in preparation). This followed the detection of the Crab pulsar by \textit{VERITAS} and \textit{MAGIC} in the very-high-energy ($>$100~GeV) regime (and now possibly up to 1~TeV; \citealp{Aleksic2011,Aleksic2012,Aliu2011}). In this paper, we will focus on the GeV band light curves of the Vela pulsar.

Despite the major advances made after nearly 50 years since the discovery of the first pulsar \citep{Hewish1968}, many questions still remain regarding the electrodynamical character of the pulsar magnetosphere, including details of the particle acceleration and pair production, current closure, and radiation of a complex multi-wavelength spectrum. Physical emission models such as the slot gap (SG; \citealp{Muslimov2003}) and outer gap (OG; \citealp{Romani1995}) fall short of explaining these global magnetospheric characteristics. More recent developments include the global magnetospheric properties. One example is the force-free (FF) inside and dissipative outside (FIDO) model \citep{Kalapotharakos2009,Kalapotharakos2014} that assumes FF electrodynamical conditions (infinite plasma conductivity, $\sigma_{\rm c}\rightarrow\infty$)  inside the light cylinder and dissipative conditions (finite $\sigma_{\rm c}$) outside. The wind models of, e.g., \citet{Petri2011} provide an alternative picture where dissipation takes place outside the light cylinder. 

Although much progress has been made using the physical models, geometric light curve modeling still presents a crucial avenue for probing the pulsar magnetosphere in the context of traditional pulsar models, as these emission geometries may be used to constrain the pulsar geometry (inclination angle $\alpha$ and the observer viewing angle $\zeta$ with respect to the spin axis $\boldsymbol\Omega$), as well as the $\gamma$-ray emission region's location and extent. This may provide vital insight into the boundary conditions and help constrain the accelerator geometry of next-generation full radiation models. Geometric light curve modeling has been performed by, e.g., \citet{Dyks2004b}, \citet{Venter2009}, \citet{Watters2009}, \citet{Johnson2014}, and \citet{Pierbattista2015} using standard pulsar emission geometries, including a two-pole caustic (TPC, of which the SG is its physical representation; \citealp{Dyks2003}), OG, and pair-starved polar cap (PC) \citep{Harding2005c} geometry. 

A notable conclusion from the 2PC was that the spectra and light curves of both the millisecond pulsar (MSP) and young pulsar populations show remarkable similarities, pointing to a common radiation mechanism and emission geometry (tied to the $B$-field structure). The assumed $B$-field structure is essential for predicting the light curves seen by the observer using geometric models, since photons are expected to be emitted tangentially to the local $B$-field lines in the corotating pulsar frame \citep{Daugherty1982}. Even a small difference in the magnetospheric structure will therefore have an impact on the light curve predictions. For all of the above geometric models, the most commonly employed $B$-field has been the retarded vacuum dipole (RVD) solution first obtained by \citet{Deutsch1955}. However, other solutions also exist. One example is the static dipole (non-rotating) field, a special case of the RVD (rotating) field~\citep{Dyks2004a}. \citet{Bai2010b} furthermore modeled high-energy (HE) light curves in the context of OG and TPC models using an FF $B$-field geometry (assuming a plasma-filled magnetosphere), proposing a separatrix layer model close to the last open field line (tangent to the light cylinder at radius $R_{\rm LC}=c/\Omega$ where the corotation speed equals the speed of light $c$, with $\Omega$ the angular speed), which extends from the stellar surface up to and beyond the light cylinder. In addition, the annular gap model of \citet{Du2010}, which assumes a static dipole field, has been successful in reproducing the main characteristics of the $\gamma$-ray light curves of three MSPs. This model does not, however, attempt to replicate the non-zero phase offsets between the $\gamma$-ray and radio profiles. 

In this paper, we investigate the impact of different magnetospheric structures (i.e., the static dipole and RVD) on the predicted $\gamma$-ray pulsar light curves. Additionally, we incorporate an offset-PC dipole $B$-field solution \citep{Harding2011a,Harding2011b} into our geometric modeling code. This analytic heuristic model mimics deviations from the static dipole such as what occur in complex solutions, e.g., dissipative (e.g., \citealp{Kalapotharakos2012,Li2012,Tchekhovskoy2013,Li2014})
 and FF (e.g., \citealp{Contopoulos1999}) fields. These complex $B$-fields usually only have numerical solutions, which are limited by the resolution of the spatial grid. Hence, it is simpler to investigate the main effects of these structures using analytical approximations such as the offset-PC dipole solution. In combination with the different $B$-field solutions mentioned above, we assume standard TPC and OG emission geometries.

\textit{Geometric} models assume constant emissivity $\epsilon_\nu$ in the rotational frame. We have also incorporated an SG $E$-field associated with the offset-PC dipole $B$-field (making this latter case an \textit{emission} model), which allows us to calculate the $\epsilon_{\nu}$ in the acceleration region in the corotating frame from first principles. We have only considered the TPC (assuming uniform $\epsilon_{\nu}$) and SG (assuming variable $\epsilon_{\nu}$ as modulated by the $E$-field) models for the offset-PC dipole $B$-field, since we do not have $E$-field expressions available for the OG model within the context of an offset-PC dipole $B$-field. The fact that we have an $E$-field solution enables us to solve the particle transport equation on each $B$-field line, yielding the particle energy (Lorentz factor $\gamma_{\rm e}$) as a function of position. We can then use this factor to test whether the particle reaches the curvature radiation reaction (CRR) limit, i.e., where acceleration balances curvature radiation losses.

We have implemented a chi-squared ($\chi^{2}$) method to search the multivariate solution space for optimal model parameters when we compare our predicted model light curves with \emph{Fermi} LAT data for the Vela pulsar. In this way, we are able to determine which $B$-field and geometric model combination yields the best light curve solution, how the different light curve predictions compare with each other, and which pulsar geometry ($\alpha$,$\zeta$) is optimal ~\citep{Breed2013,Breed2012,Breed2014}. 

In Section~\ref{sec:geometries} we describe the geometric pulsar models and $B$-field structures we considered. Section~\ref{sec:implementation} details the implementation of an offset-PC dipole $B$-field in our code. We also discuss the implementation of the associated SG $E$-field and the matching of the low-altitude and high-altitude solutions using a matching parameter (scaled radius) $\eta_{\rm c}$. We briefly describe the implementation of a $\chi^2$ method in order to find the best-fit ($\alpha$,$\zeta$) for the different model combinations. In Section~\ref{sec:results}, we present our solution of the transport equation for the offset-PC dipole $B$-field, as well as our light curve predictions and best-fit ($\alpha$,$\zeta$) contours for the Vela pulsar. In Section~\ref{sec:investigate} we investigate the effects of lowering the minimum photon energy as well as multiplying the $E$-field by a factor of 100 when constructing model light curves, before we compare, in Section~\ref{sec:comparison_models}, our results to previous multi-wavelength studies from other works. Our conclusions follow in Section~\ref{sec:conclusions}. Given the length of this paper, we summarize our main conclusions here:
\begin{enumerate}
  \item The SG $E$-field is relatively low. We thus find that the particle energy only becomes large enough to yield significant curvature radiation at large altitudes above the stellar surface, so that particles do not always attain the radiation-reaction limit. 
 \item Our overall optimal light curve fit is for the RVD field and OG model; for this $B$-field, the TPC model is statistically disfavored. 
 \item For the static dipole field, neither OG nor TPC model is significantly preferred. 
 \item We found that a smaller PC offset is favored for the offset-PC dipole field when assuming constant emissivity, and a larger PC offset favored for variable emissivity, but not significantly so. 
 \item When multiplying the SG $E$-field by a factor of 100, we found improved light curve fits, with $\alpha$ and $\zeta$ being closer to best fits from independent studies, as well as CRR at lower altitudes.
\end{enumerate}

\section{GEOMETRIC MODEL DESCRIPTION}\label{sec:geometries} 
 
\subsection{TPC and OG Geometries}\label{sub:models}

The geometric TPC pulsar model was first introduced by \citet{Dyks2003}. \citet{Muslimov2003} revived the physical SG model of \citet{Arons1983}, including general relativistic (GR) corrections, and argued that the SG model may be considered a physical representation of the TPC model. This gap geometry has a large radial extent, spanning from the neutron star (NS) surface along the last closed field line up to the light cylinder. The original definition stated that the maximum radial extent reached $R_{\rm max}{\simeq}0.8{R_{\rm LC}}$ \citep{Muslimov2004a}. This was later extended to $R_{\rm max}{\simeq}1.2{R_{\rm LC}}$ for improved fits (e.g., \citealp{Venter2009,Venter2012}). Typical transverse gap extents of $0-5\%$ of the PC angle have been used \citep{Venter2009,Watters2009}.  

The OG model was introduced by \citet{Cheng1986a,Cheng1986b} and elaborated by \citet{Romani1995}. They proposed that when the primary current passes through the neutral sheet or null-charge surface (NCS; with a radius of $R_{\rm NCS}$, i.e., the geometric surface across which the charge density changes sign) the negative charges above this sheet will escape beyond the light cylinder. A vacuum gap region is then formed (in which the $E$-field parallel to the local $B$-field, $E_{\parallel}\neq0$). Analogously, the geometric OG model has a radial extent spanning from the NCS to the light cylinder. We follow \citet{Venter2009} and \citet{Johnson2014} who considered a one-layer model with a transverse extent along the inner edge of the gap.

\subsection{$B$-field Structures}\label{sub:Bfields}

The $B$-field is one of the basic assumptions of the geometric models (others include the gap region's location and the $\epsilon_{\nu}$ profile in the gap). Several $B$-field structures have been studied in this context, including the static dipole \citep{Griffiths1995}, the RVD (a rotating vacuum magnetosphere which can in principle accelerate particles but do not contain any charges or currents; \citealp{Deutsch1955}), the FF (filled with charges and currents, but unable to accelerate particles, since the accelerating $E$-field is screened everywhere; \citealp{Contopoulos1999}), and the offset-PC dipole (which analytically mimics deviations from the static dipole near the stellar surface; \citealp{Harding2011a,Harding2011b}). A more realistic pulsar magnetosphere, i.e., a dissipative solution \citep{Kalapotharakos2012,Li2012,Tchekhovskoy2013,Li2014}, would be one that is intermediate between the RVD and the FF fields. The dissipative $B$-field is characterized by the plasma conductivity $\sigma_{\rm c}$ (e.g., \citealp{Lichnerowicz1967}), which can be chosen in order to alternate between the vacuum ($\sigma_{\rm c}\rightarrow{0}$) and FF ($\sigma_{\rm c}\rightarrow\infty$) cases (see \citealp{Li2012}).  We studied the effect of different magnetospheric structures (static dipole, RVD, and offset-PC dipole, further discussed below) and emission geometries (TPC and OG) on pulsar visibility and $\gamma$-ray pulse shape, particularly for the case of the Vela pulsar.

The solution for a $B$-field surrounding a rotating NS in vacuum was first derived by \citet[see, e.g., \citealp{Yadigaroglu1997}; \citealp{Arendt1998}; \citealp{Jackson1999}; \citealp{Cheng2000}; \citealp{Dyks2004a}]{Deutsch1955}. The general expression for this RVD field in the rotation frame (where $\hat{{\mathbf{z}}}||\boldsymbol{\Omega}$) is given by
\begin{eqnarray}\label{eq:RVD-Bfield}
{\bf{B}}_{\rm ret} & = & -\left[\frac{{\boldsymbol{\mu}}(t)}{r^3}+\frac{\dot{{\boldsymbol{\mu}}}(t)}{cr^2}+\frac{\ddot{\boldsymbol{\mu}}(t)}{c^2r}\right] \nonumber \\
& & +{\hat{{\mathbf{r}}}}{\hat{{\mathbf{ r}}}}{\boldsymbol{\cdot}}\left[3\frac{{\boldsymbol{\mu}}(t)}{r^3}+3\frac{\dot{\boldsymbol{\mu}}(t)}{cr^2}+\frac{\ddot{\boldsymbol{\mu}}(t)}{c^2r}\right],
\end{eqnarray}
with the magnetic moment
\begin{eqnarray}\label{eq:magnetic-moment}
{\boldsymbol{\mu}}(t) = \mu(\sin\alpha\cos\Omega{t}{\hat{{\mathbf{x}}}}+\sin\alpha\sin\Omega{t}{\hat{{\mathbf{y}}}}+\cos\alpha{\hat{{\mathbf{z}}}}),
\end{eqnarray}
where $\dot{\boldsymbol{\mu}}(t)$ and $\ddot{\boldsymbol{\mu}}(t)$ are the first and second time derivatives of the magnetic moment, and ($\hat{\bf{x}}$,$\hat{\bf{y}}$,$\hat{\bf{z}}$) are the Cartesian unit vectors. See \citet{Dyks2004a} for the RVD solution in spherical or Cartesian coordinates in the laboratory frame. In the limit of $r/R_{\rm LC}\rightarrow 0$, the retarded field simplifies to the non-aligned static dipole $B$-field ($\alpha\neq0$). For the static dipole the field lines are symmetric about the $\boldsymbol{\mu}$-axis, whereas the RVD is distorted due to sweepback of the field lines as the NS rotates. This has implications for the definition of the PC (see Section~\ref{sub:code}).

The offset-PC dipole is a heuristic model of a non-dipolar magnetic structure where the PCs are offset from the ${\boldsymbol{\mu}}$-axis. The $B$-field lines are therefore azimuthally asymmetric compared to those of a pure static dipole field. This leads to field lines having a smaller curvature radius $\rho_{\rm curv}$ over half of the PC (in the direction of the PC offset) compared to those of the other half. Such small distortions in the $B$-field structure are due to retardation and asymmetric currents, thereby shifting the PCs by small amounts in different directions. 

\citet{Harding2011a,Harding2011b} considered two cases, i.e., symmetric and asymmetric PC offsets. The symmetric case involves an offset of both PCs in the same direction and applies to NSs with some interior current distortions that produce multipolar components near the stellar surface. The asymmetric case is associated with asymmetric PC offsets in opposite directions and applies to PC offsets due to retardation and/or currents of the global magnetosphere. These non-dipolar $B$-field geometries are motivated by observations of thermal X-ray emission, e.g., pulse profiles from MSPs such as PSR J0437$-$4715 \citep{Bogdanov2007} and PSR J0030+0451 \citep{Bogdanov2009}, with the $B$-fields of NSs in low-mass X-ray binaries being even more distorted \citep{Lamb2009}. Magnetic fields such as the FF solution (characterized by different PC currents than those assumed in space-charge limited flow models; \citealp{Contopoulos1999,Timokhin2006}) will undergo larger sweepback of field lines near the light cylinder, and consequently display a larger offset of the PC toward the trailing side (opposite to the rotation direction) than in the RVD field (which has offset PCs due to rotation alone). In what follows, we decided to study the effect of the simpler symmetric case (which does not mimic field line sweepback of FF, RVD, or dissipative magnetospheres) on predicted light curves. In the future, one can also include the more complex asymmetric case. 

The general expression for a symmetric offset-PC dipole $B$-field in spherical coordinates $(r^{\prime},\theta^{\prime},\phi^{\prime})$ in the magnetic frame (indicated by the primed coordinates, where $\hat{{\mathbf{z}}}^{\prime}\parallel{\boldsymbol{\mu}}$) is as follows \citep{Harding2011b}
\begin{eqnarray}\label{eq:Symm-offset}
{{\mathbf{B}}}_{\rm OPCs}^{\prime} & \approx & \frac{\mu^\prime}{r^{\prime3}}\biggl[\cos\theta^{\prime}\hat{{\mathbf{r}}}^{\prime}+\frac{1}{2}(1+a)\sin\theta^{\prime}{\hat{\boldsymbol{\theta}}}^{\prime} \nonumber \\
& & -\epsilon\sin\theta^{\prime}\cos\theta^{\prime}
\sin(\phi^{\prime}-\phi_{0}^\prime)
\hat{\boldsymbol{\phi}}^{\prime}\biggr],
\end{eqnarray}
where $\mu^\prime=B_{0}R^{3}$ is the magnetic moment, $R$ the stellar radius, $B_{0}$ the surface $B$-field strength at the magnetic pole, $\phi_{0}^{\prime}$ the magnetic azimuthal angle defining the plane in which the offset occurs, and $a=\epsilon\cos(\phi^{\prime}-\phi_0^\prime)$ characterizes the offset direction in the $x^\prime-z^\prime$ plane. If $\phi^\prime_0=0$ or $\phi^\prime_0=\pi$ the offset is in the $x^\prime$ direction (i.e., along the $x^\prime$-axis). If $\phi^\prime_0=\pi/2$ or $\phi^\prime_0=3\pi/2$ the offset is in the $y^\prime$ direction. The $B$-field lines are distorted in all directions. This distortion depends on parameters $\epsilon$ (related to the magnitude of the shift of the PC from the magnetic axis) and $\phi_0^\prime$ (we choose $\phi^\prime_0=0$ in what follows). If we set $\epsilon=0$ the symmetric case reduces to a symmetric static dipole. 

The distance by which the PCs are shifted on the NS surface is given by
\begin{equation}
\Delta{r_{\rm PC}}\simeq{R\theta_{\rm PC}}\left[1-\theta_{\rm PC}^\epsilon\right],
\end{equation}
where $\theta_{\rm PC}=(\Omega{R}/c)^{1/2}$ is the standard half-angle of the PC. This effective shift of the PCs is a fraction of $\theta_{\rm PC}$; therefore it is a larger fraction of $R$ for pulsars with shorter periods \citep{Harding2011a}. \citet{Harding2011b} found that for the RVD solution, $\epsilon=0.03-0.1$, where offsets as large as $0.1$ are associated with MSPs with large $\theta_{\rm PC}$. However, $\epsilon=0.09-0.2$ is expected for FF fields, with the larger offset values related to MSPs \citep{Bai2010b}.

The difference between our offset-PC field and a dipole field that is offset with respect to the stellar center can be most clearly seen by performing a multipolar expansion of these respective fields. \citet{Lowrie2011} gives the scalar potential $W$ for an equatorially offset dipole (EOD) field as 
\begin{eqnarray}\label{eq:scalar-potential}
W^{\prime}(r^\prime,\theta^\prime)&=&\biggl[\frac{\mu^\prime}{r^{\prime2}}\cos\theta^\prime+\frac{\mu^\prime{d}}{r^{\prime3}}\sin\theta^\prime\cos\theta^\prime+\frac{\mu^\prime{d}}{2r^{\prime3}}\sin^{2}\theta^\prime\biggr], \nonumber
\end{eqnarray}
with $d$ being the offset parameter, and with the first few leading terms in $d/r^\prime$ listed above. From this potential, we may construct the magnetic field using $B = -\nabla W$:
\begin{eqnarray}\label{eq:offsetdipole}
{\bf B}_{\rm EOD}^{\prime}(r^\prime,\theta^\prime)&=&{\bf B}_{\rm dip}^{\prime}(r^\prime,\theta^\prime)+\biggl[\frac{3\mu^\prime{d}}{r^{\prime4}}\sin\theta^\prime\cos\theta^\prime+\frac{3\mu^\prime{d}}{2r^{\prime4}}\sin^{2}\theta^\prime\biggr]\hat{\mathbf{r^\prime}} \nonumber\\
& &-\biggl[\frac{\mu^\prime{d}}{r^{\prime4}}\cos^{2}\theta^\prime+\frac{\mu^\prime{d}}{r^{\prime4}}\sin^{2}\theta^\prime-\frac{\mu^\prime{d}}{r^{\prime4}}\sin\theta^\prime\cos\theta^\prime\biggr]\hat{\boldsymbol{\theta^\prime}}, \nonumber \\
&=& {\bf B}_{\rm dip}^{\prime}(r^\prime,\theta^\prime)+{\it O}\left(\frac{1}{r^{\prime4}}\right).
\end{eqnarray}
This means that an offset dipolar field may be expressed (to lowest order) as the sum of a centered dipole and two quadropolar components. Conversely, our offset-PC field may be written as
\begin{eqnarray}\label{eq:Symm-offset}
{{\mathbf{B}}}_{\rm OPCs}^{\prime}(r^\prime,\theta^\prime,\phi^\prime) & \approx & {\bf B}_{\rm dip}^{\prime}(r^\prime,\theta^\prime)+{\it O}\left(\frac{\epsilon}{r^{\prime3}}\right).
\end{eqnarray}

Therefore, we can see that the EOD consists of a centered dipole plus quadropolar and other higher-order components (Equation~(\ref{eq:offsetdipole})), while our offset-PC model (Equation~(\ref{eq:Symm-offset})) consists of a centered dipole plus terms of order $a/r^{\prime3}$ or $\epsilon/r^{\prime3}$. Since $a\sim0.2$ and $\epsilon\sim0.2$, the latter terms present perturbations (e.g., poloidal and toroidal effects) to the centered dipole. \citet{Harding2011a,Harding2011b} derived these perturbed components of the distorted magnetic field while satisfying the solenoidality condition $\nabla\cdot B=0$.

\subsection{Geometric Modeling Code}\label{sub:code}

We performed geometric light curve modeling using the code first developed by \citet{Dyks2004b}. We extended this code by implementing an offset-PC dipole $B$-field (for the symmetric case), as well as the SG $E_\parallel$-field corrected for GR effects (see Section~\ref{sec:implementation}). We solve for the PC rim as explained in Section~\ref{sub:epsilon_extend}. The shape of the PC rim depends on the $B$-field structure at $R_{\rm LC}$.  Once the PC rim has been determined, the PC is divided into self-similar (interior) rings. These rings are calculated by using open-volume coordinates ($r_{\rm ovc}$ and $l_{\rm ovc}$). After the footpoints of the field lines on a ($r_{\rm ovc}$,$l_{\rm ovc}$) grid have been determined, particles are followed along these lines in the corotating frame and emission from them is collected in bins of pulse phase $\phi_{\rm L}$ and $\zeta$, i.e., a sky map is formed by plotting the bin contents (divided by the solid angle subtended by each bin) for a given $\alpha$, and it is therefore a projection of the radiation beam. To simulate light curves, one chooses a sky map corresponding to a fixed $\alpha$, then fixes $\zeta$ and plots the intensity per solid angle.

The code takes into account the structure/geometry of the $B$-field (since the photons are emitted tangentially to the local field line), aberration of the photon emission direction (due to rotation, to first order in $r/R_{\rm LC}$), and time-of-flight delays (due to distinct emission radii) to obtain the caustic emission beam \citep{Morini1983,Dyks2004c}. However, \citet{Bai2010a} pointed out that previous studies assumed the RVD field to be valid in the instantaneously corotating frame, but actually it is valid in the laboratory frame (implying corrections that are of second order in $r/R_{\rm LC}$). This implies a revised aberration formula, which we have implemented in our code.   

\section{IMPLEMENTATION OF AN OFFSET-PC DIPOLE $B$-FIELD AND ASSOCIATED SG $E$-FIELD AND A SEARCH FOR THE BEST-FIT LIGHT CURVE}\label{sec:implementation}

\subsection{Transformation of a $B$-field from the Magnetic to the Rotational Frame}\label{sub:transform}

Since the offset-PC dipole field is specified in the magnetic frame ($\hat{\rm{\mathbf{z}}}^{\prime}\parallel\boldsymbol{\mu}^\prime$; \citealp{Harding2011b}), it is necessary to transform this solution to the (corotating) rotational frame ($\hat{\rm{\mathbf{z}}}\parallel{\boldsymbol{\Omega}}$). In order to do so, we first need to rotate the Cartesian coordinate axes specified in the rotational frame through an angle $+\alpha$ to move to the magnetic frame, and then perform transformations between the Cartesian and spherical coordinates in the magnetic frame. Only then can we transform the $B$-vector components from a spherical base unit vector set to a Cartesian one. We lastly perform a rotation of the Cartesian $B$-vector components through $-\alpha$, to move from the magnetic to the rotational frame. See the Appendix for a systematic discussion.

\subsection{Finding the PC Rim and Extending the Range of $\epsilon$}\label{sub:epsilon_extend}

The object is to find the polar angle $\theta_{*}$ at each azimuthal angle $\phi$ at the footpoints of the last open $B$-field lines, lying within a bracket $\theta_{\rm min}<\theta_{*}<\theta_{\rm max}$, such that the field line is tangent to the light cylinder. The PC rim is thus defined. The magnetic structure at the light cylinder therefore determines the PC shape \citep{Dyks2004a,Dyks2004b}. 

After initial implementation of the offset-PC dipole field in the geometric code, we discovered that we could solve for the PC rim in a similar manner as for the RVD $B$-field, but only for small values of the offset parameter $\epsilon$ ($\epsilon\lesssim{0.05-0.1}$, depending on $\alpha$). We improve the range of $\epsilon$ by varying the colatitude parameters $\theta_{\rm min}$ and $\theta_{\rm max}$, which delimit a bracket (``solution space") in colatitude thought to contain the footpoint of last open field line (tangent to the light cylinder $R_{\rm LC}$). We obtain a progressively larger range of $\epsilon$ upon decreasing $\theta_{\rm min}$ and increasing $\theta_{\rm max}$. We find a maximum $\epsilon=0.18$ valid for the full range of $\alpha$. Choosing a maximal solution bracket in colatitude would in principle work, but the code would take much longer to find the PC rim compared to when a smaller bracket (that does contain the correct solution) is used. Therefore, we generalize the search for optimal $\theta_{\rm min}$ and found (by trial and error) that the following linear equation $\theta_{\rm min}=[(-31/18)\epsilon+0.6]\theta_{\rm PC}$, for a fixed $\theta_{\rm max}=2.0$, resulted in $\theta_{\rm min}$ that yielded maximum values for $\epsilon$.

\begin{figure}
	\centering
	\includegraphics[width=16cm]{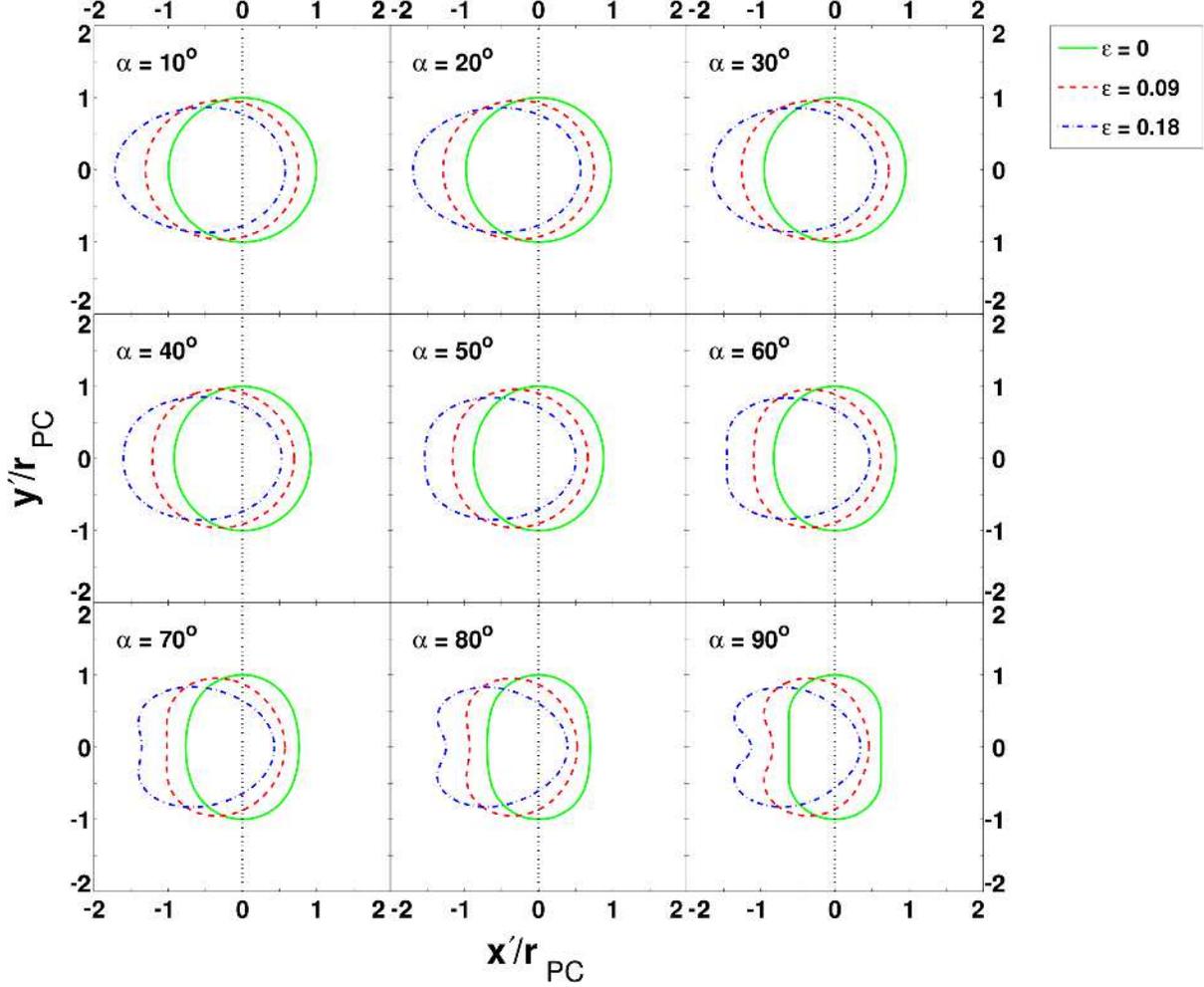}
	\caption{\label{fig:OD_PCshape} PC shapes of the offset-PC dipole $B$-field for a few cases of $\alpha$ and $\epsilon$ in the $x^\prime-y^\prime$ plane assuming that the $\boldsymbol{\mu}$-axis is located perpendicularly to the page at $(x^\prime,y^\prime)=(0,0)$ and that $\phi^\prime$ is measured counterclockwise from the positive $x^\prime$-axis. Each PC is for a different value of $\alpha$ ranging between $10^\circ$ and $90^\circ$, with $10^\circ$ resolution. For each $\alpha$ we plot the PC shape for $\epsilon$ values of $0$ (green solid circle), $0.09$ (red dashed circle), and $0.18$ (blue dashed$-$dotted circle). We note that the reference green PCs are for the static centered dipole. The horizontal line at $x^\prime=0$ (black dotted line) serves as a reference line to show the magnitude and direction of the offset as $\epsilon$ is increased.} 
\end{figure}

We illustrate the PC shape for a few cases of $\alpha$ and $\epsilon$ in Figure~\ref{fig:OD_PCshape}. We plot the PC rims ($r_{\rm ovc}=1$) in the $x^\prime-y^\prime$ plane (in the magnetic frame, in units of $r_{\rm PC}$), assuming that the $\boldsymbol{\mu}$-axis is located perpendicularly to the page at $(x^\prime,y^\prime)=(0,0)$ and that $\phi^\prime$ is measured counterclockwise from the positive $x^\prime$-axis. Each PC is for a different value of $\alpha$ in the range $10^\circ-90^\circ$ with increments of $10^\circ$. For each $\alpha$ we plot the PC shape for $\epsilon$ values of $0$ (green solid circle), $0.09$ (red dashed circle), and $0.18$ (blue dashed$-$dotted circle). The horizontal line at $x^\prime=0$ (black dotted line) serves as a reference line to show the magnitude and direction of offset as $\epsilon$ is increased. As $\alpha$ and $\epsilon$ are increased the PC shape changes considerably. For larger $\epsilon$ the PC offset is larger along the $-x^\prime$-axis (in the direction of ``unfavorably curved'' $B$-field lines). Also, as $\alpha$ increases for each $\epsilon$ the PC shape along the $x^\prime$-axis becomes narrower and irregular, e.g., compare the cases of $\epsilon=0$ and $\epsilon=0.18$ for $\alpha=90^\circ$. We note that the PC also becomes slightly narrower along the $y^\prime$-axis as $\epsilon$ increases.

\subsection{Incorporating an SG $E$-field}\label{sub:SG-Efield}

It is important to take the accelerating $E$-field into account when such expressions are available, since this will modulate the emissivity $\epsilon_\nu(s)$ (as a function of arclength $s$ along the $B$-field line) in the gap as opposed to geometric models where we assume constant $\epsilon_\nu$ per unit length in the corotating frame. For the SG case we implement the full $E$-field in the rotational frame corrected for GR effects (e.g., \citealp{Muslimov2003,Muslimov2004a}). This solution consists of a low-altitude and high-altitude limit which we have to match on each $B$-field line. The low-altitude solution is given by A. K. Harding (2016, private communication)
\begin{eqnarray}\label{eq:E_low}
E_{\parallel,{\rm low}} & \approx & {-3}{\mathcal{E}_0}\nu_{\rm SG}x^a\Big\{\frac{\kappa}{\eta^4}e_{\rm 1A}\cos\alpha+\frac{1}{4}\frac{\theta_{\rm PC}^{1+a}}{\eta}\big[e_{\rm 2A}\cos\phi_{\rm PC} \nonumber \\
 & & +\frac{1}{4}\epsilon\kappa{e_{\rm 3A}}(2\cos\phi_0^\prime-\cos(2\phi_{\rm PC}-\phi_0^\prime))\big]\sin\alpha\Big\}(1-\xi_\ast^2),
\end{eqnarray}
with $\mathcal{E}_0=(\Omega{R}/c)^2(B_r/B)B_0$, $B_r$ the radial $B$-field component, $\nu_{\rm SG}\equiv({1/4})\Delta\xi_{\rm SG}^2$, and $\Delta\xi_{\rm SG}$ the colatitudinal gap width in units of dimensionless colatitude $\xi=\theta/\theta_{\rm PC}$. Also, $x=r/R_{\rm LC}$ is the normalized radial distance in units of $R_{\rm LC}$. Here, $\kappa\approx{0.15I_{45}/R_6^3}$ is a GR compactness parameter characterizing the frame-dragging effect near the stellar surface \citep{Muslimov1997},~$I_{45}=I/10^{45}$~g~cm$^2$, $I$ the moment of inertia, $R_6=R/10^6$ cm, $\eta=r/R$ the dimensionless radial coordinate in units of $R$, $e_{\rm 1A}=1+{a}(\eta^3-1)/3$, $e_{\rm 2A}=(1+3a)\eta^{(1+a)/2}-2a$ and $e_{\rm 3A}=[({5-3a})/{\eta^{(5-a)/2}}]+2a$. The magnetic azimuthal angle $\phi_{\rm PC}$ is defined for usage with the $E$-field, being $\pi$ out of phase with $\phi^\prime$ (one chooses the negative $x$-axis toward $\boldsymbol{\Omega}$ to coincide with $\phi_{\rm PC}=0$, labeling the ``favorably curved" $B$-field lines). We define $\phi^{\prime}=\arctan(y^\prime/x^\prime)$ the magnetic azimuthal angle used when transforming the $B$-field (Section~\ref{sub:transform}). Lastly, $\xi_\ast$ is the dimensionless colatitude labeling the gap field lines (defined such that $\xi_{\ast}=0$ corresponds to the field line in the middle of the gap and $\xi_{\ast}=1$ at the boundaries; \citealp{Muslimov2003}). 

We approximate the high-altitude SG $E$-field by \citep{Muslimov2004a}
\begin{eqnarray}\label{eq:E_high}
E_{\parallel,{\rm high}} & \approx & -\frac{3}{8}\Big(\frac{\Omega{R}}{c}\Big)^3\frac{B_{\rm 0}}{f(1)}\nu_{\rm SG}x^a\Big\{\Big[1+\frac{1}{3}\kappa\Big(5-\frac{8}{\eta^3_{\rm c}}\Big)+2\frac{\eta}{\eta_{\rm LC}}\Big]\cos\alpha \nonumber \\
 & & +\frac{3}{2}\theta_{\rm PC}H(1)\sin\alpha\cos\phi_{\rm PC}\Big\}(1-\xi_\ast^2),   
\end{eqnarray}
with $f(\eta)\sim{1+0.75y+0.6y^2}$ a GR correction factor of order 1 for the dipole component of the magnetic flux through the magnetic hemisphere of radius $r$ in a Schwarzchild metric. The function $H(\eta)\sim{1-0.25y-0.16y^2-0.5(\kappa/\epsilon_{\rm g}^3)y^3(1-0.25y-0.21y^2)}$ is also a GR correction factor of order 1, with $y=\epsilon_{\rm g}/\eta$, $\epsilon_{\rm g}=r_{\rm g}/R$, and $r_{\rm g}=2GM/c^2$ the gravitational or Schwarzchild radius of the NS (with $G$ the gravitational constant and $M$ the stellar mass). The factors $f(\eta)$ and $H(\eta)$ account for the static part of the curved spacetime metric and have a value of 1 in flat space \citep{Muslimov1997}. The critical scaled radius $\eta_{\rm c}=r_{\rm c}/R$ is where the high-altitude and low-altitude $E$-field solutions are matched, with $r_{\rm c}$ the critical radius and $\eta_{\rm LC}=R_{\rm LC}/R$. This high-altitude solution (excluding the factor $x^a$) is actually valid for the SG model assuming a static (GR-corrected, non-offset) dipole field. We therefore scale the $E$-field by a factor $x^a$ to generalize this expression for the offset-PC dipole field. The general $E$-field valid from $R$ to $R_{\rm LC}$ (i.e., over the entire length of the gap) is constructed as follows (see Equation~(59) of \citealp{Muslimov2004a})
\begin{equation}\label{eq:E_total}
E_{\parallel,{\rm SG}}{\simeq}E_{\parallel,{\rm low}}\exp[-(\eta-1)/(\eta_{\rm c}-1)]+E_{\parallel,{\rm high}}.
\end{equation}
A more detailed discussion of the electrodynamics in the SG geometry may be found in \citet{Muslimov2003} and \citet{Muslimov2004a}. In the next section, we solve for $\eta_{\rm c}(P,\dot{P},\alpha,\epsilon,\xi,\phi_{\rm PC})$ where $P$ is the period and $\dot{P}$ its time derivative.

\subsection{Determining the Matching Parameter $\eta_{\rm c}$}\label{sub:matching}

At first, we matched the low-altitude and high-altitude $E$-field solutions by setting $\eta_{\rm c}=1.4$ for simplicity \citep{Breed2013}. However, we realized that $\eta_{\rm c}$ may strongly vary for the different parameters. Thus, we had to solve $\eta_{\rm c}(P,\dot{P},\alpha,\epsilon,\xi,\phi_{\rm PC})$ on each $B$-field line. In what follows we consider electrons to be the radiating particles, and our discussion will therefore generally deal with the negative of the $E$-field. Since particle orbits approximately coincide with the $B$-field lines in the corotating frame, it is important to consider the behavior of the $E$-field as a function of arclength $s$ rather than $\eta$.

\begin{figure}
	\centering
	\includegraphics[width=6cm]{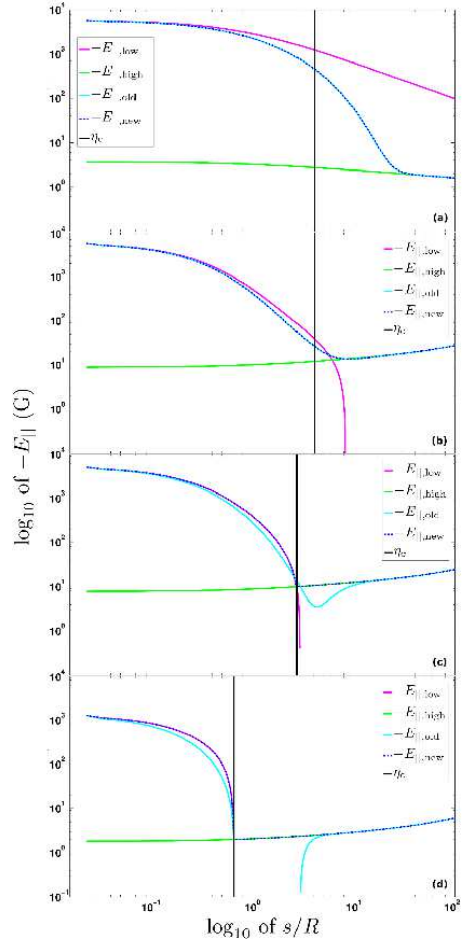}
	\caption{\label{fig:eta_C_examples} Examples of the general SG $E$-field ($E_{\parallel,\rm new}$, dashed dark blue line) we obtained by matching $E_{\parallel,\rm low}$ (magenta line) and $E_{\parallel,\rm high}$ (green line). We plotted the negative of the various $E$-fields as functions of the normalized arclength $s$ along the $B$-field lines, in units of $R$. We indicated the matching parameter $\eta_{\rm c}$ (vertical black line) by using $s_{\rm c}/R\approx\eta_{\rm c}-1$ (which is valid for low altitudes). These plots were obtained for the following parameters: $P=0.0893$~s, $B_0=1.05\times{10}^{13}$~G, $R=10^6$~cm, $M=1.4M_\odot$, $\epsilon=0.18$, and $\xi=0.975$ (i.e., $\xi_\ast=0$). In (a) we chose $\alpha=90^\circ$, and $\phi_{\rm PC}=0$. Here we use $\eta_{\rm c}=5.1$ since $\eta_{\rm cut}>\eta_{\rm LC}$. In (b) we chose $\alpha=15^\circ$, $\phi_{\rm PC}=\pi$. We find a solution of $\eta_{\rm c}=5.1$. In (c) we chose  $\alpha=30^\circ$, $\phi_{\rm PC}=\pi$. If $-E_{\parallel,\rm low}$ as well as $-E_{\parallel,\rm old}$ (as defined in Equation~(\ref{eq:E_total}), light blue line) are below $-E_{\parallel,\rm high}$ beyond some radius $\eta$, we use $\eta_{\rm cut}$ (in this case $\eta_{\rm c}=\eta_{\rm cut}=3.7$) to match $E_{\parallel,\rm low}$ and $E_{\parallel,\rm high}$, resulting in $-E_{\parallel,\rm new}$ (dashed dark blue line). In (d) we chose $\alpha=75^\circ$, $\phi_{\rm PC}=\pi$. For large $\alpha$ we observe that $-E_{\parallel,\rm low}$ changes sign over a small $\eta$ range. In this case we also use $\eta_{\rm c}=\eta_{\rm cut}=1.7$ to match the solutions.}
\end{figure}

We solve for the matching parameter in the following way. First, we calculate $E_{\parallel,\rm low}$, which is independent of $\eta_{\rm c}$, along the $B$-field. If $-E_{\parallel,\rm low}<0$ for all $\eta$, it will never intersect with $E_{\parallel,\rm high}$ and we set $\eta_{\rm c}=1.1$, thereby basically using $E_{\parallel,\rm SG}\approx E_{\parallel,\rm high}$. Second, we step through $\eta_{\rm c}$ (in the range $1.1-5.1$), calculating $E_{\parallel,\rm SG}$ and $E_{\parallel,\rm high}$ as well as the ratio $S_i=S(\eta_i)=E_{\parallel,\rm SG}(\eta_i)/E_{\parallel,\rm low}(\eta_i)$ for $i=1,..,N$ at different radii $\eta_i$. If $S_i>1$ we use $1/S_i$. We next calculate a test statistic $T(\eta_{\rm c})=\sum_i^N{(S_i-1)^2}/N$ using only $E$-field values where $-E_{\parallel,\rm low}>-E_{\parallel,\rm high}$ (i.e., we basically fit $E_{\parallel,\rm SG}$ to $E_{\parallel,\rm low}$ when $-E_{\parallel,\rm low}>-E_{\parallel,\rm high}$). We then minimize $T$ to find the optimal $\eta_{\rm c}$ (similar to what was done in Figure~2 of \citealp{Venter2009}). In Figure~\ref{fig:eta_C_examples}(a), the intersection radius $\eta_{\rm cut}>\eta_{\rm LC}$ (i.e., $E_{\parallel,\rm low}$ and $E_{\parallel,\rm high}$ do not intersect within the light cylinder) and therefore we impose the restriction that the solution of $\eta_{\rm c}$ should lie at or below 5.1. When $-E_{\parallel,\rm low}$ does not decrease as rapidly (e.g., as in Figure~\ref{fig:eta_C_examples}(b)) we find reasonable solutions. We note that $E_{\parallel,\rm SG}$ (referred to as $E_{\parallel,\rm old}$ in Figure~\ref{fig:eta_C_examples}) produces a bump when $-E_{\parallel,\rm low}$ decreases more rapidly. To circumvent this problem we test whether $-E_{\parallel,\rm SG}<-E_{\parallel,\rm high}$ and in this case we use the intersection radius $\eta_{\rm cut}$ of $E_{\parallel,\rm low}$ and $E_{\parallel,\rm high}$, rather than $\eta_{\rm c}$, to match our solutions (calling this new solution $E_{\parallel,\rm new}$; see Figure~\ref{fig:eta_C_examples}(c)). We lastly observe that for $\phi_{\rm PC}=\pi$ (on ``unfavorably curved'' field lines) for larger $\alpha$, the $-E_{\parallel,\rm low}$ field changes sign resulting in a small $\eta_{\rm c}=\eta_{\rm cut}=1.7$ value (Figure~\ref{fig:eta_C_examples}(d)). 

\begin{figure}[ht!]
    \centering
    \includegraphics[width=13cm]{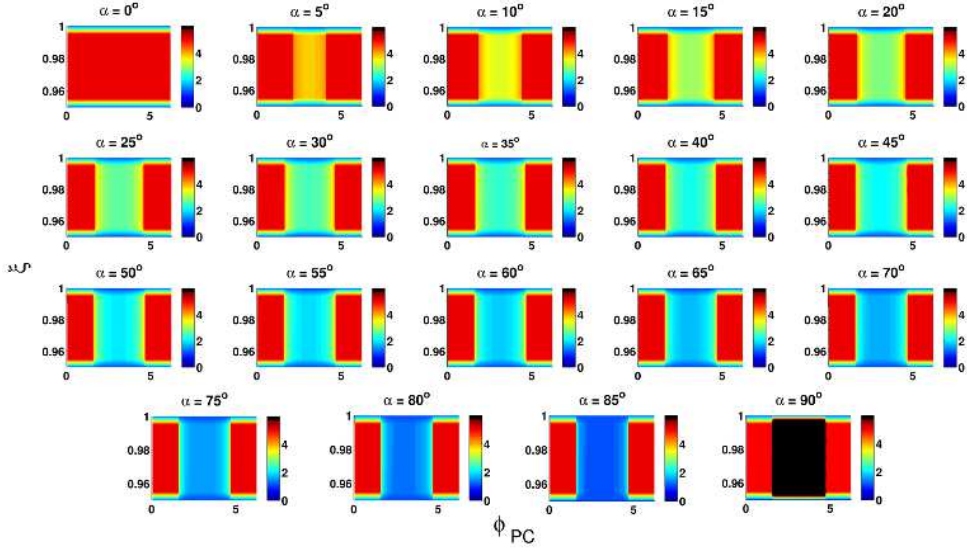}
    \caption{\label{fig:etaC_eps0} Contour plots for our solution of $\eta_{\rm c}$ for $P=0.0893$ s, $B_0=1.05\times{10}^{13}$ G, $I=0.4MR^2=1.14\times10^{45}$ g\,cm$^2$, $\alpha\in[0^\circ,90^\circ]$ with $5^\circ$ resolution, $\epsilon=0$, and $\phi_0^\prime =0$. In each case $\xi$ and $\phi_{\rm PC}$ represent the scaled colatitudinal and azimuthal magnetic coordinates, with the negative $x^\prime$-axis ($\phi_{\rm PC}=0$) directed toward the $\boldsymbol{\Omega}$-axis. The color bar represents our $\eta_{\rm c}$ solutions ranging between $1.1$ and $5.1$, with 1.0 corresponding to the NS surface and 5.1 to our limit for $\eta_{\rm c}$ when $\eta_{\rm cut}$ becomes too large. As $\alpha$ increases the second term in the $E$-field expressions starts to dominate and the solutions for $\eta_{\rm c}$ become larger for $\phi_{\rm PC}=0$ (``favorably curved'' field lines), and smaller for $\phi_{\rm PC}=\pi$ (``unfavorably curved'' field lines) until no solutions are found (e.g., the black regions where $-E_{\parallel,\rm SG}$ becomes negative).}
\end{figure}

\begin{figure}[ht!]
    \centering
    \includegraphics[width=13cm]{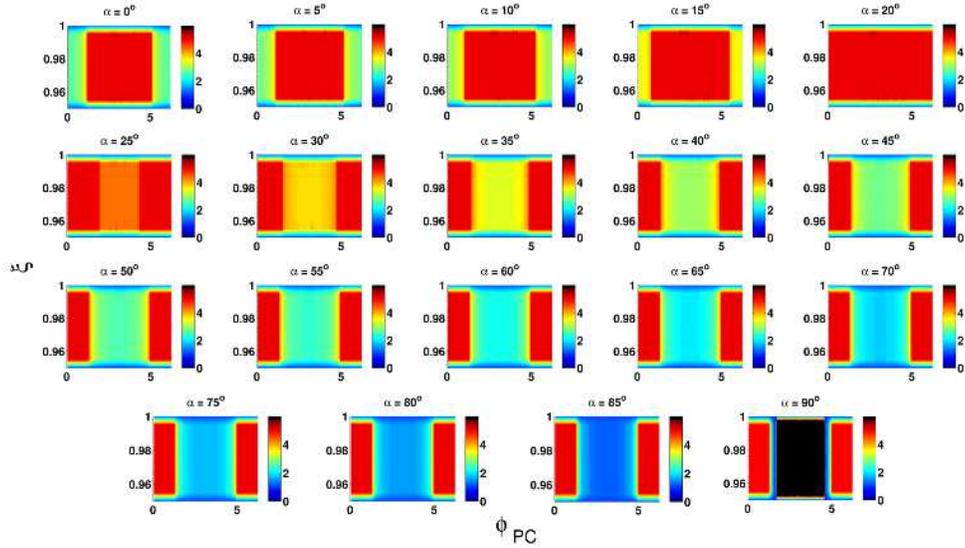}
    \caption{\label{fig:etaC_eps018} Contour plots for our solution of $\eta_{\rm c}$, similar to Figure~\ref{fig:etaC_eps0} but for $\epsilon=0.18$.}
\end{figure}

We present $\eta_{\rm c}$ contours in Figure~\ref{fig:etaC_eps0} for an offset parameter $\epsilon=0$ and in Figure~\ref{fig:etaC_eps018} for $\epsilon=0.18$. (For plotting purposes, we set $\eta_{\rm c} = 11$ when $\eta_{\rm c}>\eta_{\rm LC}$.) Since the $E$-field solutions have an $x^a=x^{\epsilon\cos(\phi^\prime-\phi^\prime_0)}=x^{-\epsilon\cos\phi_{\rm PC}}$ factor dependence, a larger (non-zero) offset results in different matching contours vs.\ the case for $\epsilon=0$. In the case of $\alpha=0$, the first term $\propto\cos\alpha$ is the only contribution to the $E$-field, with the factor $x^ae_{\rm 1A}$ (with an $\epsilon$ dependence) being initially larger at low $\eta$ for $\phi_{\rm PC}=0$ than for $\phi_{\rm PC}=\pi$ ($x^a$ dominates), but rapidly decreasing with $\eta$ ($e_{\rm 1A}$ dominates), leading to a lower value of $\eta_{\rm c}$ for $\phi_{\rm PC}=0$ (compare the first panel of Figure~\ref{fig:etaC_eps018} to the first column of Figure~\ref{fig:RRLim}). One should therefore note that the magnitude of one instance of the $E$-field with low $\eta_{\rm c}$ may initially be higher than another instance with high $\eta_{\rm c}$, but the first will decrease rapidly with $\eta$ and eventually become lower than the second. In Figure~\ref{fig:etaC_eps0} one can see that there is no $\phi_{\rm PC}$-dependence for $\alpha=0$, which is not the case for Figure~\ref{fig:etaC_eps018}. For a slightly larger $\alpha$ the second terms in Equations~(\ref{eq:E_low}) and~(\ref{eq:E_high}) start to contribute to the radiation. This is due to the $\sin\alpha$ term with an $\epsilon$ dependence that delivers an extra contribution (Figure~\ref{fig:etaC_eps018}) which is zero in the case for $\epsilon=0$ (Figure~\ref{fig:etaC_eps0}). At $\alpha=20^\circ$ the effects of the first and second terms seem to balance each other and therefore we find the same solution of $\eta_{\rm c}=5.1$ everywhere except on the SG model boundary (at $\xi\in[0.95,1.0]$) where $\eta_{\rm c}=1.1$, just as in the case of $\epsilon=0$ and $\alpha=0^\circ$. For values of $\alpha>20^\circ$ (Figure~\ref{fig:etaC_eps018}) the second term $\sim\cos\phi_{\rm PC}$ starts to dominate and thus we find solutions of $\eta_{\rm c}\sim{5.1}$ for $\phi_{\rm PC}\simeq0$ and systematically smaller solutions for $\phi_{\rm PC}\simeq\pi$ as $\alpha$ increases and the second term $\propto\sin\alpha$ becomes increasingly important (in both cases of $\epsilon$). At $\alpha=90^\circ$ we obtain the same solution as in Figure~\ref{fig:etaC_eps0} where the second term dominates (for this case $-E_{\parallel,\rm SG}<0$ for all $\eta$, since the Goldreich$-$Julian charge density $\rho_{\rm GJ}$ becomes positive). We note that the $\eta_{\rm c}$ distribution reflects two symmetries (one about $\phi_{\rm PC}=\pi$ and one about $\xi=0.975$, i.e., $\xi_{\ast}=0$, given our gap boundaries): that of the $\cos\phi_{\rm PC}$ term and that of the $(1-\xi_{\ast}^2)$ term in the $E_\parallel$ solutions. After solving for $\eta_{\rm c}$, we can solve the particle transport equation along each $B$-field line (see Section~\ref{sub:transport}).

\subsection{Chi-squared Fitting Method}\label{sec:chi2}

We apply a standard $\chi^{2}$ statistical fitting technique to assist us in objectively finding the pulsar geometry ($\alpha$,$\zeta$) which best describes the observed $\gamma$-ray light curve of the Vela pulsar. We use this $\chi^{2}$ method to determine the best-fit parameters for each of our $B$-field and geometric model combinations (spanning a large parameter space). The general expression is given by
\begin{eqnarray}\label{eq:chi2}
\chi^{2} & = & \sum_{i{\rm =1}}^{N_{{\rm bins}}}\frac{\left(Y_{{\rm d,}i}-Y_{{\rm m,}i}\right)^{2}}{\sigma_{{\rm m,}i}^{2}}\approx \sum_{i{\rm =1}}^{N_{{\rm bins}}}\frac{\left(Y_{{\rm d,}i}-Y_{{\rm m,}i}\right)^{2}}{Y_{{\rm d,}i}},
\end{eqnarray}
where $Y_{{\rm d,}i}(\phi_{{\rm L},i})$ and $Y_{{\rm m,}i}(\phi_{{\rm L},i})$ are the number of counts of the observed and modeled light curves (relative units at phase $\phi_{{\rm L},i}$), and $\sigma_{{\rm m,}i}(\phi_{{\rm L},i})$ the uncertainty of the model light curves in each phase bin $i=1,...,N_{{\rm bins}}$, with $N_{\rm bins}$ the number of bins. Since we do not know the uncertainty of the model, we approximate the model error by the data error, assuming $\sigma_{{\rm m,}i}^2(\phi_{{\rm L},i})\approx Y_{{\rm d,}i}(\phi_{{\rm L},i})$ for Poisson statistics. Since we use geometric models, with an uncertainty in the absolute {\it intensity}, we assume that the {\it shape} of the light curve is correct. The data possess a background that is also uncertain. Furthermore, \textit{Fermi} has a certain response function that influences the intrinsic shape of the light curve, which reflects the sum of counts from many pulsar rotations. Given all these uncertainties, we incorporate a free amplitude parameter $A$ to allow more freedom in terms of finding the best fit of the model light curves to the data. We normalize the model light curve to range from 0 to the maximum number of observed counts $k_{{\rm 2}}$ by using the following expression:
\begin{equation}\label{chi2_norm}
Y_{{\rm m}}^{\prime}(\phi_{{\rm L},i})=\frac{Y_{{\rm m}}(\phi_{{\rm L},i})}{(k_{1}+\epsilon_{0})}A(k_{2}-{\rm BG})+{\rm BG}\approx\frac{Y_{{\rm m}}(\phi_{{\rm L},i})}{k_{1}}k_{2},
\end{equation}
with $k_{1}={\rm max}(Y_{{\rm m}}(\phi_{{\rm L},i}))$, $k_{2}={\rm max}(Y_{{\rm d}}(\phi_{{\rm L},i}))$, $\epsilon_{0}$ a small value added to ensure that we do not divide by zero, $A$ a free normalization parameter, and BG the background level of $Y_{{\rm d}}(\phi_{{\rm L},i})$. We treat the data as being cyclic so we need to ensure that the model light curve is cyclic as well. The model light curve has to be re-binned in order to have the same number of bins in $\phi_{\rm L}$ as the data \citep{Abdo2013}. We use a Gaussian Kernel Density Estimator function to rebin and smooth the model light curve \citep{Parzen1962}. Furthermore, we also introduce the free parameter $\Delta\phi_{\rm L}$ which represents an arbitrary phase shift of the model light curve so as to align the model and data peaks. We choose the phase shift $\Delta\phi_{\rm L}$ as a free parameter due to the uncertainty in the definition of $\phi_{\rm L}=0$ (see, e.g., \citealp{Johnson2014} who also used $A$ and $\Delta\phi_{\rm L}$). Importantly, we note that we have not changed the relative position (the radio-to-$\gamma$ phase lag $\delta$), since this is a crucial model prediction. The radio and $\gamma$-ray emission regions are tied to the same underlying $B$-field structure, and $\delta$ therefore reflects important physical conditions (or model assumptions) such as a difference in emission heights of the radio and $\gamma$-ray beams.

After preparation of the model light curve, we search for the best-fit solution for each of our $B$-field and gap combinations over a parameter space of $\alpha\in[0^{\circ},90^{\circ}]$, $\zeta\in[0^{\circ},90^{\circ}]$ (both with $1^{\circ}$ resolution), $0.5<A<1.5$ with $0.1$ resolution, and $0<\Delta\phi_{\rm L}<1$ with $0.05$ resolution. For a chosen $B$-field and model geometry we iterate over each set of parameters and search for a local minimum $\chi^{2}$ value at a particular $\alpha$ and $\zeta$. Once we have iterated over the entire parameter space ($\alpha$,$\zeta$,$A$,$\Delta\phi_{\rm L}$), we obtain a global minimum value for $\chi^{2}$ (also called the optimal $\chi^{2}$):
\begin{eqnarray}\label{eq:chi2-opt}
\chi_{{\rm opt}}^{2} & \approx & \sum_{i=1}^{N_{{\rm bins}}}\frac{\left(Y_{{\rm d,}i}-Y_{{\rm opt,}i}\right)^{2}}{Y_{{\rm d,}i}}.
\end{eqnarray}

If faint pulsars are modeled, Poisson statistics will be sufficient to describe the observations. For the bright Vela, however, we assume Gaussian statistics which yields small errors, since the emission characteristics are more significant than those of faint pulsars. However, these small errors on the data yield large values for the reduced optimal $\chi^2$ value $\chi_{{\rm opt}}^{2}/N_{\rm dof}\gg 1$. We therefore need to rescale (to compensate for the uncertainty in $\sigma_{{\rm m,}i}$) the $\chi^{2}$ values by $\chi_{{\rm opt}}^{2}$ and multiply by the number of degrees of freedom $N_{\rm dof}$ (the difference between $N_{\rm bins}$ and number of free parameters). The scaled $\chi^{2}$ is presented by \citep{Pierbattista2015}:
\begin{eqnarray}\label{eq:scaled-chi2}
\xi^{2} & = & N_{{\rm dof}}\frac{\chi^{2}}{\chi_{{\rm opt}}^{2}}.
\end{eqnarray}
From Equation~(\ref{eq:scaled-chi2}) the $\xi^{2}$ for the optimal model are as follows
\begin{eqnarray}\label{eq:scaled-chi2-opt}
\xi_{{\rm opt}}^{2} & = & N_{{\rm dof}}\frac{\chi_{{\rm opt}}^{2}}{\chi_{{\rm opt}}^{2}}=N_{{\rm dof}},
\end{eqnarray}
with $\xi_{\rm opt}^{2}/N_{\rm dof}=\xi_{\rm opt,\nu}^{2}=1$ the reduced $\xi_{\rm opt}^{2}$.  

If one wishes to compare the optimal model to alternative models, e.g., in our case a $B$-field combined with several geometric models, confidence contours for $68\%$ ($1\sigma$), $95.4\%$ ($2\sigma$), and $99.73\%$ ($3\sigma$) can be constructed by estimating the difference in the $\xi_{{\rm opt}}^{2}$ and the $\xi^{2}$ of the alternative models:
\begin{eqnarray}\label{eq:difference-chi2}
\Delta\xi^{2} & =\xi^{2}-\xi_{{\rm opt}}^{2}=N_{{\rm dof}}\left(\chi^{2}/\chi_{{\rm opt}}^{2}-1\right).\end{eqnarray}
The confidence intervals can be estimated by reading the $\Delta\xi^{2}$ (i.e., $\Delta\xi_{1\sigma,\mu_{{\rm dof}}}^{2}$, $\Delta\xi_{2\sigma,\mu_{{\rm dof}}}^{2}$, and $\Delta\xi_{3\sigma,\mu_{{\rm dof}}}^{2}$) values from a standard $\chi^{2}$ table for the specified confidence interval at $\mu_{{\rm dof}}=2$ (corresponding to the two-dimensional ($\alpha,\zeta$) grid \citealp{Lampton1976}).  Using these values for $\Delta\xi^{2}$ and $\xi_{{\rm opt}}^{2}=N_{{\rm dof}}$, we can determine $\xi^{2}=\xi^2_{\rm opt}+\Delta\xi^2=N_{\rm dof}+\Delta\xi^2$ (i.e., $\xi_{1\sigma}^{2}$, $\xi_{2\sigma}^{2}$, and $\xi_{3\sigma}^{2}$) from Equation~(\ref{eq:difference-chi2}), which is the value at which we plot each confidence contour. To enhance the contrast of the colors on the filled $\chi^{2}$ contours, we plot ${\rm log_{10}\xi^{2}}$ on an ($\alpha$,$\zeta$) grid, with a minimum value of ${\rm log_{10}}\xi_{{\rm opt}}^{2}={\rm log_{10}}(N_{{\rm dof}})=1.98$ (corresponding to the best-fit solution by construction, i.e., after rescaling, with $N_{\rm dof}=100-4=96$ in our study). The best-fit solution is therefore positioned at $\xi_{{\rm opt}}^{2}=96$ and enclosed by the confidence contours with values of $\xi_{1\sigma,\mu_{{\rm dof}}}^{2}=96+2.30$, $\xi_{2\sigma,\mu_{{\rm dof}}}^{2}=96+6.17$, and $\xi_{3\sigma,\mu_{{\rm dof}}}^{2}=96+11.8$ (see Equation~(\ref{eq:difference-chi2}); \citealp{Press1992}). We determine errors on $\alpha$ and $\zeta$ for the best-fit solution of each $B$-field and model combination using the $3\sigma$ interval connected contours. We choose errors of $1^{\circ}$ for cases when the errors were smaller than one degree (given a model resolution of $1^{\circ}$). See Section~\ref{sub:contours_LCs}. For the TPC and RVD model combination, we encountered poor qualitative and statistical fits using the $\chi^{2}$ method, thus an alternative solution had to be selected even though the $\chi^2$ value was larger.

\section{RESULTS}\label{sec:results}

\subsection{Solving the Particle Transport Equation}\label{sub:transport}

Once we solved $\eta_{\rm c}$ (see Section~\ref{sub:matching}), we could calculate the general $E$-field ($E_{\parallel,{\rm new}}$) in order to solve the particle transport equation (in the corotating frame) to obtain the particle energy $\gamma_{\rm e}(\eta,\phi,\xi_{\ast})$, necessary for determining the CR emissivity \citep{Breed2013}
\begin{equation}
\dot{\gamma}=\dot{\gamma}_{\rm gain}+\dot{\gamma}_{\rm loss}=\frac{eE_{\parallel,{\rm new}}}{m_{\rm e}c}-\frac{2e^2\gamma_{\rm e}^4}{3\rho^2_{\rm curv}m_{\rm e}c}=\frac{1}{m_{\rm e}c^2}\left[{ceE_{\parallel,{\rm new}}}-\frac{2ce^2\gamma_{\rm e}^4}{3\rho^2_{\rm curv}}\right],
\end{equation}
with $\dot{\gamma}_{\rm gain}$ the gain (acceleration) rate, $\dot{\gamma}_{\rm loss}$ the loss rate, $e$ the electron charge, $m_{\rm e}$ the electron mass, and $m_{\rm e}c^2$ the rest-mass energy; CRR (taking only CR losses into account) occurs when the energy gain balances the losses and $\dot{\gamma}=0$. 

\begin{figure}
	\centering
	\includegraphics[width=15cm]{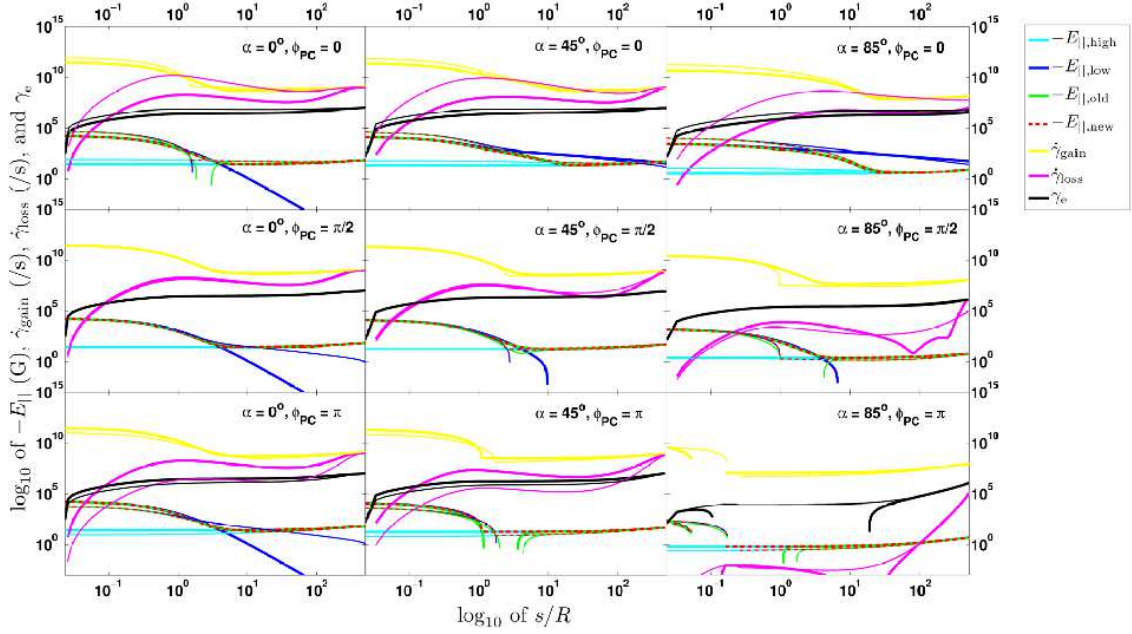}
	\caption{\label{fig:RRLim} Plot of $\log_{10}$ of $-E_{\parallel,{\rm high}}$ (solid cyan line), $-E_{\parallel,{\rm low}}$ (solid blue line), the general $-E_{\parallel,{\rm SG}}$-field (using $\eta_{\rm c}$ as the matching parameter; $-E_{\parallel, {\rm old}}$, solid green line) and a corrected $E$-field in cases where a bump was formed using the standard matching procedure (i.e., setting $\eta_{\rm c}=\eta_{\rm cut}$; $-E_{\parallel, {\rm new}}$, dashed red line), gain rate $\dot{\gamma}_{\rm gain}$ (solid yellow line), loss rate $\dot{\gamma}_{\rm loss}$ (solid magenta line), and the Lorentz factor $\gamma_{\rm e}$ (solid black line) as a function of arclength $s/R$. In each case we used $P=0.0893$~s, $B_0=1.05\times{10}^{13}$ G (corrected for GR effects), $I=0.4MR^2=1.14\times10^{45}$ g\,cm$^2$, and $\xi=0.975$ (i.e., $\xi_{\ast}=0$). On each panel we represent the curves for $\epsilon=0$ (thick lines) and $\epsilon=0.18$ (thin lines). The first column is for $\alpha=0^\circ$, the middle one for $\alpha=45^\circ$, and the third one for $\alpha=85^\circ$. For each column, the top panel is for ``favorably curved" field lines ($\phi_{\rm PC}=0$), the middle panel for $\phi_{\rm PC}=\pi/2$, and the bottom panel for ``unfavorably curved" field lines ($\phi_{\rm PC}=\pi$). These choices reflect the values of $\phi_{\rm PC}$ at the stellar surface; they may change as the particle moves along the $B$-field line, since $B_{\phi}\neq 0$.}
\end{figure}

In Figure~\ref{fig:RRLim} we plot the $\log_{10}$ of $-E_{\parallel,{\rm high}}$ (solid cyan line), $-E_{\parallel,{\rm low}}$ (solid blue line), the general $-E_{\parallel,{\rm SG}}$ field (using $\eta_{\rm c}$ as the matching parameter; $-E_{\parallel, {\rm old}}$, solid green line) and a corrected $E$-field in cases where a bump was formed using the standard matching procedure (see Section~\ref{sub:matching}, i.e., setting $\eta_{\rm c}=\eta_{\rm cut}$; $-E_{\parallel, {\rm new}}$, dashed red line), $\dot{\gamma}_{\rm gain}$ (solid yellow line), $\dot{\gamma}_{\rm loss}$ (solid magenta line), and $\gamma_{\rm e}$ (solid black line) as a function of arclength $s/R$ along the $B$-field line. For each case we show $\epsilon=0$ (thick lines) and $\epsilon=0.18$ (thin lines) on the same plot. We note that the values for $\phi_{\rm PC}$ appearing on the figure are actually values on the stellar surface; these may change along the $B$-field lines. In the first column we set $\alpha=0^\circ$ so that only the first term~$\propto\cos\alpha$ in Equations~(\ref{eq:E_low}) and~(\ref{eq:E_high}) contributes. In this case, $-E_{\parallel,{\rm low}}\propto{x^a}[1+a(\eta^3-1)/3]$, where $a=-\epsilon\cos\phi_{\rm PC}$. The sign of $x^a$ stays the same but at small values of $\eta$, $x^a\approx{3}$ for $\phi_{\rm PC}=0$ (top panel) and $x^a\approx{1/3}$ for $\phi_{\rm PC}=\pi$ (bottom panel). This explains why $-E_{\parallel,{\rm low}}$ is higher for $\phi_{\rm PC}=0$ than for $\phi_{\rm PC}=\pi$ (for $\epsilon\neq 0$) at low $\eta$. In the case of $\phi_{\rm PC}=\pi/2$ (middle panel), $a=0$ and the values of $E_{\parallel,{\rm low}}$ are very nearly the same for both $\epsilon=0$ and $\epsilon\neq 0$, the only difference stemming from the $B$-field structure that enters into Equation~(\ref{eq:E_low}) through $\mathcal{E}_0$. The low-altitude $E$-field is therefore enhanced in the direction of the ``favorably curved'' $B$-field lines ($\phi_{\rm PC}=0$). The term $[1+a(\eta^3-1)/3]$ changes sign beyond some $\eta$ for $\phi_{\rm PC}=0$, explaining the behavior of $-E_{\parallel,{\rm low}}$ beyond $s\approx{2R}$. This effect is also noticeable in Figure~\ref{fig:etaC_eps018} for $\alpha\approx 0^\circ$, where smaller values of the matching altitude $\eta_{\rm c}$ are found for $\phi_{\rm PC}=0$ or $\phi_{\rm PC}=2\pi$. This is contrary to the case of $\phi_{\rm PC}=\pi$ (``unfavorably curved'' $B$-field lines) where the sign stays the same for all $\eta$. Correspondingly, $\eta_{\rm c}\approx 5$ around $\phi_{\rm PC}\neq 0$ in Figure~\ref{fig:etaC_eps018} for $\alpha\approx 0^\circ$. For $-E_{\parallel,{\rm high}}\propto{x^a[1+2\eta/\eta_{\rm LC}]}$, we note that for $\phi_{\rm PC}=(0,\pi/2,\pi)$, $x^a$ changes from the following low to high values: $(3\rightarrow 1,1\rightarrow 1,1/3\rightarrow 1)$ (for $\epsilon\neq 0$, otherwise $x^a=1$) while $[1+2\eta/\eta_{\rm LC}]$ changes from $1\rightarrow 3$ (the number to the left of $\rightarrow$ is valid for $\eta\ll 1$ and the number to the right of $\rightarrow$ is valid for $\eta\gg 1$). The combined effect of these two terms is such that $-E_{\parallel,{\rm high}}\propto(3\rightarrow 3,1\rightarrow 3,1/3\rightarrow 3)$ for $\phi_{\rm PC}=(0,\pi/2,\pi)$ and $\epsilon\neq 0$, and $-E_{\parallel,{\rm high}}\propto\rightarrow 3$ for $\epsilon=0$, for all $\phi_{\rm PC}$. Therefore the high-altitude $E$-field is again enhanced in the direction of the ``favorably curved'' $B$-field lines at low values of $\eta$ (and suppressed for ``unfavorably curved'' $B$-field lines), coinciding at high $\eta$ for the different $\phi_{\rm PC}$. It follows that $\dot{\gamma}_{\rm gain}$, $\gamma_{\rm e}$, and $\dot{\gamma}_{\rm loss}$ are higher for $\phi_{\rm PC}=0$, and lower for $\phi_{\rm PC}=\pi$ ($\epsilon\neq 0$). We note that $E_{\parallel,{\rm high}}$ is not so strongly dependent on $\eta$ as $E_{\parallel,{\rm low}}$. There is no effect in changing $\phi_{\rm PC}$ in the case of $\alpha=0^\circ$ and $\epsilon=0$, since the first term does not depend on $\phi_{\rm PC}$. 

In the second and third columns of Figure~\ref{fig:RRLim} we set $\alpha=45^\circ$ and $\alpha=85^\circ$ respectively. We note that $-E_{\parallel,{\rm high}}$ displays the same behavior at low $\eta$ as previously: for $\phi_{\rm PC}=0$, $-E_{\parallel,{\rm high}}^{\epsilon\neq 0}>-E_{\parallel,{\rm high}}^{\epsilon=0}$; these are nearly equal for $\phi_{\rm PC}=\pi/2$, and $-E_{\parallel,{\rm high}}^{\epsilon\neq 0}<-E_{\parallel,{\rm high}}^{\epsilon=0}$ for $\phi_{\rm PC}=\pi$. For $\phi_{\rm PC}=\pi/2$, the first term of $-E_{\parallel,{\rm high}}$ dominates the second, and for $\phi_{\rm PC}=\pi$, the second term of $-E_{\parallel,{\rm high}}$ is always negative, but the positive first term dominates and therefore $-E_{\parallel,{\rm high}}$ does not change sign as $\eta$ increases. A similar behavior is also seen for $-E_{\parallel,{\rm low}}$ (boosted for non-zero $\epsilon$ and $\phi_{\rm PC}=0$). For $\alpha=45^\circ$ (second column of Figure~\ref{fig:RRLim}), the second term $\propto\sin\alpha$ now contributes, stopping $-E_{\parallel,{\rm low}}$ from changing sign along $\eta$ for $\phi_{\rm PC}=0$ (vs.\ the case of $\alpha=0$, first column, first row). The second term of $-E_{\parallel,{\rm low}}\sim x^a\cos\phi_{\rm PC}$ is comparable to the first at low $\eta$, but quickly dominates as $\eta$ increases for $\phi_{\rm PC}=0$. The second term of $-E_{\parallel,{\rm low}}$ remains positive so that we find $\eta_{\rm c}=5.1$ in this case (compare Figures~\ref{fig:etaC_eps0} and~\ref{fig:etaC_eps018}). For $\phi_{\rm PC}=\pi/2$ we note that $-E_{\parallel,{\rm low}}^{\epsilon\neq 0}$ becomes negative with $\eta$. For $\phi_{\rm PC}=\pi$, the second term of $-E_{\parallel,{\rm low}}$ is negative, forcing this field to change sign; this change takes slightly longer to occur when $\epsilon\neq 0$. The fact that $-E_{\parallel,{\rm high}}$ is positive leads to a ``recovery" of the total $E$-field, so that it becomes positive again at larger $\eta$. The effect of matching the $E$-field is seen in the evolution of $\gamma_{\rm e}(s)$ since $\gamma_{\rm e}$ is determined by $E_\parallel$.

In the third column of Figure~\ref{fig:RRLim}, we indicate the $E_\parallel$-field\footnote[4]{As mentioned above, the sign of $\rho_{\rm GJ}$ changes near $\alpha\approx 90^\circ$ and $\phi_{\rm PC}\approx\pi$, and we ignored such $B$-field lines when calculating the emission from the pulsar. This is also the reason why our plots of $\gamma_{\rm e}$ vs.\ $s/R$ only go up to $\alpha=85^\circ$, since we only consider electrons to be emitters of HE $\gamma$-rays in our model.} for $\alpha=85^\circ$. For $\phi_{\rm PC}=0$, the behaviors of $-E_\parallel$, $\gamma_{\rm e}$, and $\dot{\gamma}$ are very similar to  the first and second columns in Figure~\ref{fig:RRLim}. For $\phi_{\rm PC}=\pi/2$ (i.e., $a=0$, as in the case for $\epsilon=0$) the first and second terms are positive for all $\eta$ and comparable in magnitude for $-E_{\parallel,{\rm low}}$ so that the sign change in the case of $\epsilon\neq 0$ may be ascribed to the structure of the offset-PC dipole $B$-field ($\phi_{\rm PC}$ increases along the field line since $B_\phi\neq0$ so that $\cos\phi_{\rm PC}$ becomes negative). For $\phi_{\rm PC}=\pi$, $-E_{\parallel,{\rm low}}^{\epsilon\neq 0}$ is not smaller than $-E_{\parallel,{\rm low}}^{\epsilon=0}$ at low $\eta$. In this case the second term of $-E_{\parallel,{\rm low}}$ is always negative, and the sum of the two terms are very close for both $\epsilon\neq 0$ and $\epsilon=0$ so that $-E_{\parallel,{\rm low}}^{\epsilon\neq 0}\approx-E_{\parallel,{\rm low}}^{\epsilon=0}$. Again, we see that $-E_{\parallel,{\rm low}}$ becomes negative, but the positive $-E_{\parallel,{\rm high}}$ leads to a recovery.

We lastly notice that the CRR limit is in fact reached in some cases, but only at high altitudes (the yellow and magenta lines reach the same value): e.g., beyond $\eta\approx{0.7R_{\rm LC}}/R$ for $\phi_{\rm PC}=0$ and $\alpha=0^\circ$, and beyond $\eta\approx{R_{\rm LC}}$ for $\phi_{\rm PC}=0$ and $\alpha=45^\circ$. The notable exception occurs at large $\alpha$ where the first term of the $E$-field becomes lower and the second term plays a larger role, leading to smaller gain rates and therefore smaller Lorentz factors. We note the importance of actually solving $\eta_{\rm c}(P,\dot{P},\alpha,\epsilon,\xi,\phi_{\rm PC})$ on each $B$-field line. Previously we set $\eta_{\rm c}=1.4$ for all cases and found that the particles did not attain the CRR limit \citep{Breed2013}. Only when we allowed larger values of $\eta_{\rm c}$ was $-E_{\parallel,{\rm low}}$ boosted and we found particles reaching the CRR limit in many more cases. The relatively low SG $E$-field leads to small caustics on the phase plots constructed for photon energies $>100$~MeV (see Section~\ref{sub:PPs_LCs}). We therefore anticipate that a higher $E$-field should lead to CRR being reached at lower altitudes, as well as to extended caustic structures on these phase plots, resulting in qualitatively different light curve shapes (see Section~\ref{sub:100Epar}). 

\subsection{Phase plots and Light Curves}\label{sub:PPs_LCs}

We next perform simulations using a geometric modeling code (Section~\ref{sub:code}) which has the following free parameters: $\alpha$, $\zeta$, and $\epsilon$ (in case of the offset-PC dipole field). We fix the scaled co-latitude of the innermost ring of the gap ($r_{\rm ovc}^{\rm min}=0.95$), PC rim ($r_{\rm ovc}^{\rm max}=1.00$), and gap width $w=r_{\rm ovc}^{\rm max}-r_{\rm ovc}^{\rm min} = 0.05$. We choose $\alpha\in[0^{\circ},90^{\circ}]$ with a $1^{\circ}$ resolution, since the simulations show symmetry in both the northern and southern hemispheres of the pulsar (as per assumption). This implies that the emission signature for $\alpha$ is the same as to that of $\pi-\alpha$, except that the phase is shifted by half a rotation, i.e., $\phi_{\rm L}=0.5$. This model symmetry is also visible in $\zeta$ (the radiation pattern is a mirror image about $\zeta=90^\circ$, also including a phase shift of $\Delta\phi_{\rm L}=0.5$). For the offset-PC dipole we choose for the TPC (assuming uniform $\epsilon_{\nu}$) and SG (assuming variable $\epsilon_{\nu}$) models a range of $\epsilon\in[0.00,0.18]$ with a resolution of $0.03$. Our sky maps use ranges of $\phi_{\rm L}\in[-0.5,0.5]$ and $\zeta\in[0.5^{\circ},179.5^{\circ}]$, both with a resolution of $2^{\circ}$. Unique emission characteristics are visible in the light curves depending on the choice of $\alpha$, $\zeta$, $B$-field structure, and emission geometry \citep{Dyks2004b,Seyffert2015}. 

\begin{figure}
	\centering
	\includegraphics[width=12cm]{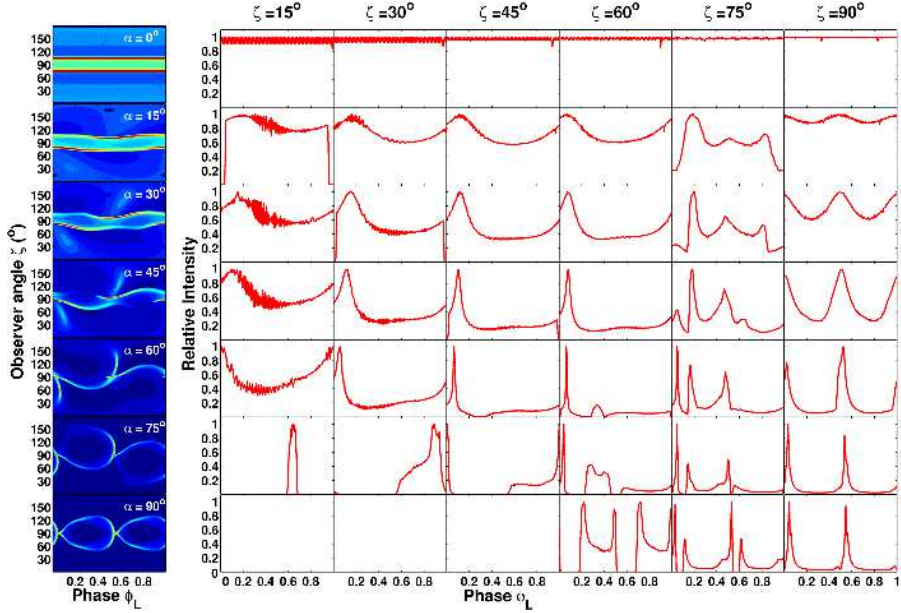}
	\caption{\label{fig:OffsetEps000} Phase plots (first column) and light curves (second column and onward) for the TPC model assuming an offset-PC dipole field, for a fixed value of $\epsilon=0.00$ and constant $\epsilon_{\nu}$. Each phase plot is for a different $\alpha$ value ranging from $0^{\circ}$ to $90^{\circ}$ with a $15^{\circ}$ resolution, and their corresponding light curves are denoted by the solid red lines for different $\zeta$ values, ranging from $15^{\circ}$ to $90^{\circ}$, with a $15^{\circ}$ resolution.}
\end{figure}

In Figure~\ref{fig:OffsetEps000} we present the phase plots and light curves we obtained for the offset-PC dipole $B$-field and TPC model combination, for $\epsilon=0$ (equivalent to the static dipole solution). For larger values of $\alpha$ the caustics extend over a larger range in $\zeta$, with the emission forming a ``closed loop,'' which is a feature of the static dipole $B$-field at $\alpha=90^{\circ}$. The TPC model is visible at nearly all angle combinations, since some emission occurs below the NCS for this model, in contrast to the OG model. However, for $\alpha=90^{\circ}$ and $\zeta$ below $45^{\circ}$ no light curves are visible, i.e., no emission is observed due to the ``closed loop'' structure of the caustics. The TPC light curves exhibit relatively more off-pulse emission than the OG ones. In the TPC model, emission is visible from both magnetic poles, forming double peaks in some cases, whereas in the OG model emission is visible from a single pole. One does obtain double peaks in the OG case, however, when the line of sight crosses the caustic at two different phases. If we compare Figure~\ref{fig:OffsetEps000} with the phase plots and light curves for the static dipole case and the TPC model, we notice that they are identical at all angle combinations. This important test case implies that we successfully transformed our offset-PC dipole $B$-field, as discussed in the Appendix. In Figure~\ref{fig:OffsetEps018} we chose an offset parameter $\epsilon=0.18$ assuming constant $\epsilon_\nu$. If we compare Figures~\ref{fig:OffsetEps000} and~\ref{fig:OffsetEps018}, we notice that this larger offset $\epsilon$ results in qualitatively different phase plots and light curves, e.g., modulation at small $\alpha$. Also, the caustics occupy a slightly larger region of phase space and seem more pronounced for larger $\epsilon$ and $\alpha$ values, accompanied by the same change in the position of the PCs as in Figure~\ref{fig:OffsetEps000}. The light curve shapes are also slightly different.
 
In Figure~\ref{fig:OffsetEps000E} we present phase plots and light curves for the offset-PC dipole $B$-field and $\epsilon=0$, obtaining a variable $\epsilon_\nu(s)$ due to using an SG $E$-field solution (with CR the dominating process for emitting $\gamma$-rays; see Sections~\ref{sub:SG-Efield}). The caustic structure and resulting light curves are qualitatively different for various $\alpha$ compared to the constant $\epsilon_\nu$ case. The caustics appear smaller and less pronounced for larger $\alpha$ values (since $E_\parallel$ becomes lower as $\alpha$ increases), and extend over a smaller range in $\zeta$. In Figure~\ref{fig:OffsetEps018E} we chose $\epsilon=0.18$, finding a variable $\epsilon_\nu$. If we compare Figure~\ref{fig:OffsetEps018E} with Figure~\ref{fig:OffsetEps000E} we note a new emission structure close to the PCs for small values of $\alpha$ and $\zeta\approx(0^\circ,180^\circ)$. This reflects the boosted $E_\parallel$-field on the ``favorably curved'' $B$-field lines (with $E_\parallel\propto{x^a}\cos\alpha$, with $a=-\epsilon{\cos\phi_{\rm PC}}$ and $\phi_{\rm PC}=0$; see Figure~\ref{fig:RRLim}). In Figure~\ref{fig:OffsetEps018E} there is also more phase space filled than in Figure~\ref{fig:OffsetEps000E}. The light curves generally display only one broad peak with less off-peak emission compared to Figure~\ref{fig:OffsetEps000}. As $\alpha$ and $\zeta$ increase, more peaks become visible, with emission still visible from both poles as seen for larger $\alpha$ and $\zeta$ values, e.g., $\alpha=75^{\circ}$ and $\zeta=75^{\circ}$.

If we compare Figure~\ref{fig:OffsetEps000} with Figure~\ref{fig:OffsetEps000E} (also Figure~\ref{fig:OffsetEps018} with Figure~\ref{fig:OffsetEps018E}), we notice that when we take $E_{\parallel}$ into account the phase plots and light curves change considerably. For example, for $\alpha=90^{\circ}$ in the constant $\epsilon_\nu$ case, a ``closed loop'' emission pattern is visible in the phase plot, which is different compared to the small ``wing-like'' emission pattern in the variable $\epsilon_\nu$ case. \textit{Therefore, we see that both the $B$-field and $E$-field have an impact on the predicted light curves.} This small ``wing-like'' caustic pattern is due to the fact that we only included photons in the phase plot with energies $>100$~MeV. Given the relatively low $E$-field there are only a few photons with energies exceeding $100$~MeV. 

To further illustrate the effect of changing $\epsilon$, we present phase plots for $\alpha=70^{\circ}$ in Figure~\ref{fig:a70z50}(a) and their corresponding light curves at $\zeta=50^{\circ}$ associated with the particular phase plot in Figure~\ref{fig:a70z50}(b), using an offset-PC dipole field and TPC model, with $\epsilon$ ranging from 0.00 to 0.18 with intervals of 0.03, and assuming constant $\epsilon_\nu$. The caustic structure is slightly different for different values of $\epsilon$. For $\epsilon=0$ the light curve has a single peak and as $\epsilon$ increases, the peak becomes slightly narrower. Also, for larger $\epsilon$ values, the caustic structure becomes slightly broader and more pronounced. Figures~\ref{fig:a70z50}(c) and~\ref{fig:a70z50}(d) represent the offset-PC dipole field and SG model with variable $\epsilon_{{\rm \nu}}$. When we compare the phase plots of Figures~\ref{fig:a70z50}(a) and~\ref{fig:a70z50}(c), the caustics are dimmer and smaller due to the low SG $E$-field, and the light curves display less off-peak emission. As $\epsilon$ increases, some small features become more pronounced. 

\subsection{$\chi^2(\alpha,\zeta)$ Contours and Best-fit Light Curves}\label{sub:contours_LCs}

In this section we present our best-fit solutions of the simulated light curves using the Vela data from {\it Fermi}. We plot some example contours of ${\rm log_{10}\xi^{2}}$ (color bar) as well as the optimal ($\alpha$,$\zeta$) combination. We determine errors on $\alpha$ and $\zeta$ for the optimal solution of each $B$-field and gap model combination using a bounding box delimited by a minimum and maximum value in both $\alpha$ and $\zeta$ which surrounds the $3\sigma$ contour. We illustrate this in Figure~\ref{fig:bestcontours}(a), with the white lines indicating the bounding box $[\alpha_{{\rm min}},\alpha_{{\rm max}}]$ and $[\zeta_{{\rm min}},\zeta_{{\rm max}}]$ (see enlargement in bottom left corner of panel~(a)). We choose errors of $1^{\circ}$ for cases when the errors were smaller than $1^\circ$ (given our chosen resolution of $1^{\circ}$). In Figure~\ref{fig:bestcontours}(a) we indicate by a white star our overall best statistical fit for an OG model using an RVD field at $\alpha=78{_{-1}^{+1}}^{\circ}$, $\zeta=69{_{-1}^{+2}}^{\circ}$, $A=1.3$, and $\Delta\phi_{\rm L}=0$ by a white star. The curved region ranging from $\zeta=70^{\circ}$ downward to $\zeta=10^{\circ}$, over the entire range of $\alpha$, is caused by the caustic structure as seen in the phase plots (i.e., there is no emission visible for low values of $\alpha$ and $\zeta$ -- the turquoise bottom-left region). 

The corresponding light curve fit of the model (solid red line) for the best-fit geometry to the Vela data (blue histogram) is also shown (Figure~\ref{fig:bestcontours}(b)). The observed light curve represents weighted counts per bin as a function of normalized phase $\phi_{\rm L}=[0,1]$ \citep{Abdo2013}. The model light curve yields a qualitatively good fit to the Vela data, exhibiting distinct qualitative features including the three peaks at the same phases, with roughly the same width, as seen in the Vela data. The OG model fits the data qualitatively better than the TPC model since the OG model displays less off-pulse emission, as seen in the phase plots and light curves in Section~\ref{sub:PPs_LCs}. 

In Figure~\ref{fig:bestcontours}(c) we present our significance contour ${\rm log_{10}\xi^{2}}$ and in Figure~\ref{fig:bestcontours}(d) the corresponding best-fit light curve for a TPC model assuming an offset-PC dipole field, with $\epsilon=0.00$ and a constant $\epsilon_{\nu}$. We find an optimal solution at $\alpha=73{_{-2}^{+3}}^{\circ}$, $\zeta=45{_{-4}^{+4}}^{\circ}$, $A=1.3$, and $\Delta\phi_{\rm L}=0.55$. Disconnected confidence intervals are visible in this case, with the $3\sigma$ errors (using only the small connected confidence contour) on $\alpha$ and $\zeta$ yielding larger values than in Figure~\ref{fig:bestcontours}(a). The best-fit model light curve yields a less satisfactory fit to the Vela data, although the model exhibits one peak coinciding with the second peak in the data. The model also displays a low level of off-peak emission similar to the data. 

In Figure~\ref{fig:bestcontours}(e) we present our significance contour ${\rm log_{10}\xi^{2}}$ and in Figure~\ref{fig:bestcontours}(f) the corresponding best-fit light curve for an SG model using an offset-PC dipole field, with $\epsilon=0.15$ and a variable $\epsilon_{\nu}$. For this combination, we find a best-fit solution at $\alpha=76{_{-1}^{+3}}^{\circ}$, $\zeta=48{_{-11}^{+15}}^{\circ}$, $A=0.7$, and $\Delta\phi_{\rm L}=0.55$. The model light curve yields a reasonable fit to the Vela data, but the peaks are low (constrained by the low level of off-peak emission, i.e., the $\chi^2$ prefers a small value for $A$), with the first peak being very broad and a small bump preceding the second peak when compared to the data.

\section{FURTHER INVESTIGATION}\label{sec:investigate}

\subsection{Light Curves in a Different Waveband}\label{sub:1MeV}

Since the SG $E$-field (see Section~\ref{sub:SG-Efield}) is low, CRR is reached in most cases but only at high $\eta$ and small $\alpha$. This low $E$-field also causes the phase plots to display small caustics which result in ``missing structure." Therefore, we investigate the effect on the light curves of the offset-PC dipole $B$-field and SG model combination when we \textit{lower the minimum photon energy} $E_{\rm min}$ from 100 to 1~MeV, above which we construct phase plots. In the CRR limit we can determine the CR cutoff of the CR photon spectrum as follows, using the formula of \citet{Venter2010}
\begin{equation}\label{ECRcut}
E_{\rm CR}\sim{4}E_{\parallel,\rm 4}^{3/4}\rho_{\rm curv, 8}^{1/2} \quad {\rm GeV},
\end{equation}
with $\rho_{\rm curv, 8}\sim\rho_{\rm curv}/{10^8}$ cm the curvature radius of the $B$-field line and $E_{\parallel,4}\sim E_{\parallel}/{10^4}$ statvolt cm$^{-1}$, the $E$-field parallel to the $B$-field. For our given SG $E$-field with a magnitude of $E_\parallel\sim10^2$~statvolt~cm$^{-1}$ the estimated cutoff is $E_{\rm CR}\sim90$~MeV. This leads to pulsar emission being emitted in the hard X-ray waveband, and cannot be compared via $\chi^2$ to \textit{Fermi} ($>100$~MeV) data for the Vela pulsar. As an illustration, we present the phase plots and light curves in Figure~\ref{fig:OffsetEps018MeV} for $\epsilon=0.18$ and $E_{\rm min}>1$~MeV. If we compare Figure~\ref{fig:OffsetEps018MeV} with Figure~\ref{fig:OffsetEps018E} we notice that a larger region of phase space is filled by caustics, especially at larger $\alpha$, e.g., at $\alpha=90^\circ$ the visibility is enhanced. The peaks are also wider at low $\alpha$. Sometimes extra emission features appear, leading to small changes in the light curve shapes.  

\subsection{Effect of Increasing the $E$-field}\label{sub:100Epar}

Additionally, we investigate what the effect is on the light curves when we \textit{increase the $E$-field.} As a test we multiply the $E$-field by a factor of 100. Using Equation~(\ref{ECRcut}) we estimate a cutoff energy $E_{\rm CR}\sim4$~GeV which is in the energy range of \textit{Fermi} ($>100$~MeV). We present the phase plots and light curves for this larger $E$-field in Figure~\ref{fig:OffsetEps018GeV} for the offset-PC dipole and SG model solution with $\epsilon=0.18$. If we compare Figure~\ref{fig:OffsetEps018GeV} with Figure~\ref{fig:OffsetEps018E} we notice that more phase space is filled by caustics, especially at larger $\alpha$. At $\alpha=90^\circ$ the visibility is again enhanced. The caustic structure becomes wider and more pronounced, with extra emission features arising as seen at larger $\alpha$ and $\zeta$ values. This leads to small changes in the light curve shapes. At smaller $\alpha$ values, the emission around the PC forms a circular pattern that becomes smaller as $\alpha$ increases. These rings around the PCs become visible since the low $E$-field is boosted, leading to an increase in bridge emission as well as higher signal-to-noise ratio. At low $\alpha$ the background becomes feature-rich, but not at significant intensities, however.

When we boost the low $E$-field, we find that the CRR limit is in fact reached almost immediately at lower $\eta$ for certain parameter combinations of $\alpha$, and $\phi_{\rm PC}$, as shown in Figure~\ref{fig:RRLim_100Epar}. We also obtained a better $\chi^2$ best-fit solution for this larger $E$-field compared to the usual one, for $\epsilon=0.00$ at $\alpha=75^{+3}_{-1}$, $\zeta=51^{+2}_{-5}$, $A=1.1$, and $\Delta\phi_{\rm L}=0.55$. In Figure~\ref{fig:bestcontourLC_Epar} we show our significance contour ${\rm log_{10}\xi^{2}}$ on the left and the corresponding best-fit light curve on the right. This offset-PC dipole $B$-field and SG model (using the increased $E$-field) combination therefore provides an overall optimal fit, second only to the RVD and OG model combination (see Section~\ref{sec:comparison_models}).

\section{COMPARISON OF BEST-FIT PARAMETERS FOR DIFFERENT MODELS}\label{sec:comparison_models}

We next follow the same approach as \citet{Pierbattista2015} to compare the various optimal solutions of the different models, in two ways: (i) per $B$-field and model combination and (ii) overall (for all $B$-field and model combinations). We determine the difference between the scaled\footnote[5]{We therefore first scale the $\chi^2$ values using the optimal value obtained for a particular $B$-field, and second we scale these using the overall optimal value irrespective of $B$-field.} $\chi^2$ of the optimal model, $\xi_{\rm opt}^{2}$, and the other models ($\xi^{2}$) using Equation~(\ref{eq:difference-chi2}), substituting $N_{\rm dof}=96$, as summarized in Table~\ref{Summary}. The best-fit parameters for each $B$-field and geometric model combination, including the case for $100E_\parallel$, are summarized in Table~\ref{Summary}. The table includes the different model combinations, the optimal unscaled $\chi^{2}$ value for each combination, the best-fit free parameters with $3\sigma$ errors on $\alpha$ and $\zeta$, and the comparison between models per $B$-field ($\Delta\xi^2_{\rm B}$) and overall ($\Delta\xi^2_{\rm all}$, with $\Delta\xi^{2}=0$ representing the best-fit solution for each $B$-field or the overall optimal fit; \citealp{Pierbattista2015}). We also include several multi-wavelength independent fits (all for the Vela pulsar). 

In Figure~\ref{fig:ModelComparisonB} we label the different $B$-field structures assumed in the various models as well as the overall comparison along the $x$-axis, and plot $\Delta\xi^{2}_{\rm B}$ and $\Delta\xi^{2}_{\rm all}$ on the $y$-axis. We represent the TPC geometry with a circle, the OG with a square, and for the offset-PC dipole field we represent the various $\epsilon$ values for constant $\epsilon_{\nu}$ by different colored stars, for variable $\epsilon_{\nu}$ by different colored left pointing triangles, and for the case of $100E_\parallel$ by different colored upright triangles, as indicated in the legend. The dashed horizontal lines indicate the confidence levels we obtained by calculating the expected $\Delta\xi^{2}$ values using an online $\chi^{2}$ statistical calculator\footnote[6]{http://easycalculation.com/statistics/critical-value-for-chi-square.php} for $N_{\rm dof}=96$ degrees of freedom,\footnote[7]{We note that \citet{Pierbattista2015} assumed that $\Delta\xi^2$ follows a $\chi^2$ distribution with $N_{\rm dof}$ degrees of freedom. We will follow this approximation here, assuming that the best-fit model provides a good fit to the observed light curves. The degrees of freedom may in reality slightly differ, however, and the matter is complicated by the fact that we want to statically compare non-nested models. A Monte Carlo approach would be preferable to find these significance levels. However, our main conclusions will not change for slight changes in these levels (which may be different for each $B$-field and model combination), and so we do not pursue this matter any further.} i.e., using $p$-values of $p_{1\sigma}=1-0.682$, $p_{2\sigma}=1-0.954$, and $p_{3\sigma}=1-0.9973$. We found critical values of $\Delta\xi^{2}=102.06$ ($1\sigma$), $120.60$ ($2\sigma$), and $139.05$ ($3\sigma$) respectively. These confidence levels are used as indicators of when to reject or accept an alternative fit compared to the optimum fit. The last column represents fits for all models, irrespective of $B$-field.

For the static dipole field the TPC model gives the optimum fit and the OG model lies within $1\sigma$, implying that the OG geometry may provide an acceptable alternative fit to the data in this case. For the RVD field the TPC model is significantly rejected beyond the $3\sigma$ level (not shown on plot), and the OG model is preferred. We show three cases for the offset-PC dipole field, including the TPC model assuming constant $\epsilon_{\nu}$, the SG model assuming variable $\epsilon_{\nu}$, and the latter with an $E_\parallel$-field increased by a factor of $100$. The optimal fits for the offset-PC dipole field and TPC model reveal that a smaller offset ($\epsilon$) is generally preferred for constant $\epsilon_{\nu}$, while a larger offset is preferred for variable $\epsilon_{\nu}$ (but not significantly), with all alternative fits falling within $1\sigma$. However, when we  increase $E_\parallel$, a smaller offset is again preferred for the SG and variable $\epsilon_\nu$ case. When we compare all model and $B$-field combinations with the overall best fit (i.e., rescaling the $\chi^2$ values of all combinations using the optimal fit involving the RVD $B$-field and OG model), we notice that the static dipole and TPC model falls within $2\sigma$, whereas the static OG model lies within $3\sigma$. We also note that the usual offset-PC dipole $B$-field and TPC model combination (for all $\epsilon$ values) is above $1\sigma$ (with some fits $<2\sigma$), but the offset-PC dipole $B$-field and SG model combination (for all $\epsilon$ values) is significantly rejected ($>3\sigma$). However, the case of the offset-PC dipole field and a higher SG $E_\parallel$ for all $\epsilon$ values leads to a recovery, since all the fits fall within $2\sigma$ and delivers an overall optimal fit for $\epsilon=0$, second only to the RVD and OG model fit. 

Several multi-wavelength studies have been performed for Vela, using the radio, X-ray, and $\gamma$-ray data, in order to find constraints on $\alpha$ and $\zeta$. We only fit the $\gamma$-ray light curve, because we did not want to bias our results by using a geometric radio emission model \citep{DeCesar2013}. However, \citet{Johnston2005} determined the radio polarization position angle from polarization data by applying a rotating vector model (RVM) fit to the datafinding best-fit values of $\alpha=53^{\circ}$ and $\zeta=59.5^{\circ}$, with an impact angle of $\beta=\zeta-\alpha=6.5^{\circ}$. \citet{Ng2008} applied a torus-fitting technique \citep{Ng2004} to fit the \textit{Chandra} data in order to constrain the Vela X pulsar wind nebula (PWN) geometry, deriving a value of $\zeta=63.6{_{-0.05}^{+0.07}}^{\circ}$ represented by the dashed black line in Figure~\ref{fig:Comparison_alphazeta}. \citet{Watters2009} modeled light curves using the PC, TPC, and OG geometries in conjunction with an RVD field, thereby constraining the geometrical parameters $\alpha$, $\zeta$, and also finding small $\beta$ in the case of the PC model. They found a good fit for their TPC model at $\alpha=62^{\circ}-68^{\circ}$ and $\zeta=64^{\circ}$, and for the OG geometry at $\alpha=75{}^{\circ}$ and $\zeta=64^{\circ}$. We find that our best-fit values for the RVD field, for both the TPC and OG models, are in good agreement with those found by \citet{Watters2009}. \citet{DeCesar2013} followed a similar approach to ours, but for the RVD and FF $B$-fields combined with emission geometries such as the SG (symmetric and asymmetric cases) and OG. They have different free model parameters including $\alpha$, $\zeta$, $w$ (gap width), and $R_{\rm max}$ (maximum emission radius), and determined errors on their best fits using the $3\sigma$ confidence intervals. They found best-fit solutions for the RVD and OG model at $\alpha=88{_{-3}^{+2}}^{\circ}$ and $\zeta=66.5{_{-1}^{+1}}^{\circ}$, which is within $10^{\circ}$ or less compared to our best-fit solution. Their overall best fit was for the FF $B$-field and OG geometry, with $\alpha={80^{+1}_{-1}}^{\circ}$ and $\zeta={53^{+1}_{-1}}^{\circ}$. \citet{Pierbattista2015} found a best-fit solution for Vela using the RVD field and OG model combination at $\alpha={71_{-2}^{+2}}^{\circ}$ and $\zeta=83{_{-2}^{+2}}^{\circ}$, with $\zeta$ exceeding the best-fit solution we found by nearly $15^{\circ}$. However, they fit both the $\gamma$-ray and radio light curves, which may explain this discrepancy. We summarize all these multi-wavelength fits and more in Table~\ref{Summary}.

We graphically summarize the best-fit $\alpha$ and $\zeta$, with errors, from this and other works in Figure~\ref{fig:Comparison_alphazeta}. We notice that the best fits generally prefer a large $\alpha$ or $\zeta$ or both. It is encouraging that many of the best-fit solutions lie near the $\zeta$ inferred from the PWN torus fitting \citep{Ng2008}, notably for the RVD $B$-field. A significant fraction of fits furthermore lie near the $\alpha-\zeta$ diagonal, i.e., they prefer a small impact angle, probably due to radio visibility constraints \citep{Johnson2014}. For an isotropic distribution of pulsar viewing angles, one expects $\zeta$ values to be distributed as $\sin(\zeta)$ between $\zeta=[0^\circ,90^\circ]$, i.e., large $\zeta$ values are much more likely than small $\zeta$ values, which seems to correspond to the large best-fit $\zeta$ values we obtain. There seems to be a reasonable correspondence between our results obtained for geometric models and those of other authors, but less so for the offset-PC dipole $B$-field, and in particular for the SG $E$-field case. The lone fit near $(20^\circ,70^\circ)$ may be explained by the fact that a very similar fit, but one with slightly worse $\chi^2$, is found at $(50^\circ,80^\circ)$. If we discard the non-optimal TPC / SG fits, we see that the optimal fits will cluster near the other fits at large $\alpha$ and $\zeta$. Although our best fits for the offset-PC dipole $B$-field are clustered, it seems that increasing $\epsilon$ leads to a marginal decrease in $\zeta$ for the TPC model (light green) and opposite for SG (dark green), but not significantly (see Table~\ref{Summary}). For our increased SG $E$-field case (brown) we note that the fits now cluster inside the gray area above the fits for the static dipole and TPC, and offset-PC dipole for both the TPC and SG geometries. 

\section{CONCLUSIONS}\label{sec:conclusions}

We investigated the impact of different magnetospheric structures on predicted $\gamma$-ray pulsar light curve characteristics. We extended our code, which already included the static dipole and RVD $B$-fields, by implementing an additional $B$-field, i.e., the symmetric offset-PC dipole field \citep{Harding2011a,Harding2011b} characterized by an offset $\epsilon$ of the magnetic PCs. We also included the full accelerating SG $E$-field corrected for GR effects up to high altitudes. For the offset-PC dipole field we only considered the TPC (assuming uniform $\epsilon_{\nu}$) and SG (modulating the $\epsilon_{\nu}$ using the $E$-field) models, since we do not have $E$-field expressions available for the OG model for this particular $B$-field. We matched the low-altitude and high-altitude solutions of the SG $E_\parallel$ by determining the matching parameter $\eta_{\rm c}(P,\dot{P},\alpha,\epsilon,\xi,\phi_{\rm PC})$ on each field line in multivariate space. Once we calculated the general $E$-field we could solve the particle transport equation. This yielded the particle energy $\gamma_{\rm e}(\eta)$, necessary for determining the CR $\epsilon_\nu$ and to test whether the CRR limit is attained. For the case of a variable $\epsilon_\nu$, we found that the CRR limit is reached for many parameter combinations (of $\alpha, \epsilon$ and $\phi_{\rm PC}$; see Figure~\ref{fig:RRLim}), albeit only at large $\eta$. A notable exception occurred at large $\alpha$ where the first term of each $E$-field expression (e.g., Equations~(\ref{eq:E_low}) and~(\ref{eq:E_high})) became lower and the second term played a larger role, leading to smaller gain rates and therefore smaller Lorentz factors $\gamma_{\rm e}$. 

We concluded that the magnetospheric structure and emission geometry have an important effect on the predicted $\gamma$-ray pulsar light curves. \textit{However, the presence of an $E$-field may have an even greater effect than small changes in the $B$-field and emission geometries:} When we included an SG $E$-field, thereby modulating $\epsilon_\nu$, the resulting phase plots and light curves became qualitatively different compared to the geometric case.       

We fit our model light curves to the observed {\it Fermi}-measured Vela light curve for each $B$-field and geometric model combination. We found that the RVD field and OG model combination fit the observed light curve the best for $(\alpha,\zeta,A,\Delta\phi_{\rm L})=({78_{-1}^{+1}}^\circ,{69_{-1}^{+2}}^\circ,1.3,0.00)$ and an unscaled $\chi^2=3.84\times{10}^4$. As seen in Figure~\ref{fig:ModelComparisonB}, for the RVD field an OG model is significantly preferred over the TPC model, given the characteristically low off-peak emission. For the other field and model combinations there was no significantly preferred model (per $B$-field), since all the alternative models may provide an acceptable alternative fit to the data, within $1\sigma$. The offset-PC dipole field for constant $\epsilon_{\nu}$ favored smaller values of $\epsilon$, and for variable $\epsilon_{\nu}$ larger $\epsilon$ values, but not significantly so ($<1\sigma$). When comparing all cases (i.e., all $B$-fields), we noted that the offset-PC dipole field for variable $\epsilon_{\nu}$ was significantly rejected ($>3\sigma$). 

We further investigated the effect which the SG $E_\parallel$ had on our predicted light curves in two ways. First, we lowered the minimum photon energy from $E_{\rm min}=100$~MeV to $E_{\rm min}=1$~MeV, leading to emission in the hard X-ray waveband. We noted new caustic structures and emission features on the resulting phase plots and light curves that were absent when $E_{\rm min}>100$~MeV. Since we wanted to compare our model light curves to \textit{Fermi} data we increased the usual low SG $E$-field by a factor of 100 (with a spectral cutoff $E_{\rm CR}\sim4$~GeV). When solving the particle transport equation, we noticed that the CRR limit is now reached in most cases at lower $\eta$. The increased $E$-field also had a great impact on the phase plots, e.g., extended caustic structures and new emission features as well as different light curve shapes emerged. We also compared the best-fit light curves for the offset-PC dipole $B$-field and $100E_\parallel$ combination for each $\epsilon$ (Figure~\ref{fig:ModelComparisonB}) and noted that a smaller $\epsilon$ was again preferred (although not significantly; $<1\sigma$). However, when we compared this case to the other $B$-field and model combinations, we found statistically better $\chi^2$ fits for all $\epsilon$ values with an optimal fit at $\alpha={75^{+3}_{-1}}^{\circ}$, $\zeta={51^{+2}_{-5}}^{\circ}$, $A=1.1$, and $\Delta\phi_{\rm L}=0.55$ for $\epsilon=0$, being second in quality only to the RVD and OG model fit. 

We graphically compared the best-fit $\alpha$ and $\zeta$, with errors, from this and other works in Figure~\ref{fig:Comparison_alphazeta}. We noted that many of the best-fit solutions cluster inside the gray area at larger $\alpha$ and $\zeta$. Some fits lie near the $\alpha-\zeta$ diagonal (possibly due to radio visibility constraints in some cases) as well as near the $\zeta$ inferred from the PWN torus fitting \citep{Ng2008}, notably for the RVD $B$-field. There was reasonable correspondence between our results obtained for geometric models and those of other independent studies. When we discarded the non-optimal TPC / SG fits, we saw that the optimal fits clustered near the other fits at large $\alpha$ and $\zeta$. For our increased SG $E$-field and offset-PC dipole combination (brown) we noted that these fits clustered at larger $\alpha$ and $\zeta$.  

There have been several indications that \textit{the SG $E$-field may be larger than initially thought.} For example, (i) population synthesis studies found that the SG $\gamma$-ray luminosity may be too low, pointing to an increased $E$-field and / or particle current through the gap \citep[e.g.,][]{Pierbattista2015}. Furthermore, if the $E$-field is too low, one is not able to reproduce the (ii) observed spectral cutoffs of a few GeV (Section~\ref{sub:1MeV}; \citealp{Abdo2013}). We found additional indications for an enhanced SG $E$-field. A larger $E$-field (increased by a factor of 100) led to (iii) statistically improved $\chi^2$ fits with respect to the light curves. Moreover, the inferred best-fit $\alpha$ and $\zeta$ parameters for this $E$-field (iv) clustered near the best fits of independent studies. We additionally observed that a larger SG $E$-field also (v) increased the particle energy gain rates and therefore yielded a larger particle energy $\gamma_{\rm e}$ (giving CR that is visible in the \textit{Fermi} band) as well as leading to a CRR regime already close to the stellar surface. These evidences may point to a reconsideration of the boundary conditions assumed by \citet{Muslimov2004a} which suppressed the $E_{\parallel}$ at high altitudes. They assumed equipotentiality of the SG boundaries as well as the steady state drift of charged particles across the SG $B$-field lines, implying $E_\perp\approx0$ at high-altitudes, with the flux of charges remaining constant up to high altitudes. One possible way to bring self-consistency may be implementation of the newly developed FIDO model that includes global magnetospheric properties and calculates the $B$-field and $E$-field self-consistently.

We envision several future projects that may emanate from this study. One could continue to extend the range of $\epsilon$ for which our code finds the PC rim, since more complex field solutions, e.g., the dissipative and FF field structures, may be associated with larger PC offsets. However, the offset-PC dipole solutions have limited applicability to outer magnetosphere emission since they use the static dipole frame and do not model the field line sweep back. Therefore, it would be preferable to investigate the $B$-fields and $E$-fields of the dissipative models and solve the transport equation to test if the particles reach the CRR limit. The effect of these new fields on the phase plots and light curves can also be studied. There is also potential for multi-wavelength studies, such as light curve modeling in the other energy bands, e.g., combining radio and $\gamma$-ray light curves (see \citealp{Seyffert2010,Seyffert2012,Pierbattista2015}). One could furthermore model energy-dependent light curves, such as those available for Vela and other bright pulsars using {\it Fermi} data (e.g., \citealp{Abdo2009}). Lastly, model phase-resolved spectra can be constructed which is an important test of the $E_\parallel$-field magnitude and spatial dependence.

\acknowledgements
We thank Marco Pierbattista, Tyrel Johnson, Lucas Guillemot, and Bertie Seyffert for fruitful discussions. This work is based on the research supported wholly / in part by the National Research Foundation of South Africa (NRF; Grant Numbers 87613, 90822, 92860, 93278, and 99072). The Grantholder acknowledges that opinions, findings and conclusions or recommendations expressed in any publication generated by the NRF supported research is that of the author(s), and that the NRF accepts no liability whatsoever in this regard. A.K.H.\ acknowledges the support from the NASA Astrophysics Theory Program. C.V.\ and A.K.H.\ acknowledge support from the \textit{Fermi} Guest Investigator Program. 

\section*{APPENDIX \\ TRANSFORMATION OF THE $B$-FIELD FROM THE MAGNETIC TO THE ROTATIONAL FRAME}\label{AppA}
Consider a general $B$-field specified in the magnetic frame (indicated by primed coordinates), in terms of spherical coordinates
\begin{equation}\label{eq:Bsphere}
{\mathbf{B}}^{\prime}(r^{\prime},\theta^{\prime},\phi^{\prime})=B_{r}^{\prime}(r^{\prime},\theta^{\prime},\phi^{\prime})\hat{\rm{\mathbf{r}}}^{\prime}+B_{\theta}^{\prime}(r^{\prime},\theta^{\prime},\phi^{\prime})\hat{\boldsymbol{\theta}}^{\prime}+B_{\phi}^{\prime}(r^{\prime},\theta^{\prime},\phi^{\prime})\hat{\boldsymbol{\phi}}^{\prime}.
\end{equation}
The spherical unit vectors may be expressed in terms of Cartesian unit vectors \citep[e.g.,][]{Griffiths1995}:
\begin{equation}\label{eq:unitvect}
\begin{array}{ccrcrcr}
{\hat{\rm {\mathbf{r}}}}^{\prime} & = & (\sin\theta^{\prime}\cos\phi^{\prime}){\hat{\rm {\mathbf{x}}}}^{\prime} & + & (\sin\theta^{\prime}\sin\phi^{\prime}){\hat{\rm {\mathbf{y}}}}^{\prime} & + & (\cos\theta^{\prime}){\hat{\rm {\mathbf{z}}}}^{\prime},\\
{\hat{\rm {\boldsymbol{\theta}}}}^{\prime} & = & (\cos\theta^{\prime}\cos\phi^{\prime}){\hat{\rm {\mathbf{x}}}}^{\prime} & + & (\cos\theta^{\prime}\sin\phi^{\prime}){\hat{\rm {\mathbf{y}}}}^{\prime} & - & (\sin\theta^{\prime}){\hat{\rm {\mathbf{z}}}}^{\prime},\\
{\hat{\rm {\boldsymbol{\phi}}}}^{\prime} & = & -(\sin\phi^{\prime}){\hat{\rm {\mathbf{x}}}}^{\prime} & + & (\cos\phi^{\prime}){\hat{\rm {\mathbf{y}}}}^{\prime}. &
\end{array}
\end{equation}
To transform the $B$-vector components from a spherical base unit vector set to a Cartesian one, substitute Equation~(\ref{eq:unitvect}) into (\ref{eq:Bsphere}) and rearrange according to unit basis vectors and compare with
\begin{equation}\label{eq:xyzprime}
{\mathbf{B}}^{\prime}(r^{\prime},\theta^{\prime},\phi^{\prime})=B_{x}^{\prime}(r^{\prime},\theta^{\prime},\phi^{\prime})\hat{\rm{\mathbf{x}}}^{\prime}+B_{y}^{\prime}(r^{\prime},\theta^{\prime},\phi^{\prime})\hat{\rm{\mathbf{y}}}^{\prime}+B_{z}^{\prime}(r^{\prime},\theta^{\prime},\phi^{\prime})\hat{\rm{\mathbf{ z}}}^{\prime}.
\end{equation}
This yields
\begin{equation}\label{eq:Bprime}
\begin{array}{ccrcrcr}
B_{x}^{\prime}(r^{\prime},\theta^{\prime},\phi^{\prime}) & = & B_{r}^{\prime}\sin\theta^{\prime}\cos\phi^{\prime} & + & B_{\theta}^{\prime}\cos\theta^{\prime}\cos\phi^{\prime} & - & B_{\phi}^{\prime}\sin\phi^{\prime},\\
B_{y}^{\prime}(r^{\prime},\theta^{\prime},\phi^{\prime}) & = & B_{r}^{\prime}\sin\theta^{\prime}\sin\phi^{\prime} & + & B_{\theta}^{\prime}\cos\theta^{\prime}\sin\phi^{\prime} & + & B_{\phi}^{\prime}\cos\phi^{\prime},\\
B_{z}^{\prime}(r^{\prime},\theta^{\prime},\phi^{\prime}) & = & B_{r}^{\prime}\cos\theta^{\prime} & - & B_{\theta}^{\prime}\sin\theta^{\prime}. &
\end{array}
\end{equation}
To transform the spherical coordinates in Equation~(\ref{eq:Bprime}) to Cartesian coordinates in the magnetic frame, use
\begin{equation}\label{eq:coordinates}
\begin{array}{ccl}
x^{\prime} & = & r^{\prime}\sin\theta^{\prime}\cos\phi^{\prime},\\
y^{\prime} & = & r^{\prime}\sin\theta^{\prime}\sin\phi^{\prime},\\
z^{\prime} & = & r^{\prime}\cos\theta^{\prime}.
\end{array}
\end{equation}
We then obtain
\begin{equation}\label{eq:xyzprime_new}
{\mathbf{B}}^{\prime}(x^{\prime},y^{\prime},z^{\prime})=B_{x}^{\prime}(x^{\prime},y^{\prime},z^{\prime})\hat{\rm{\mathbf{x}}}^{\prime}+B_{y}^{\prime}(x^{\prime},y^{\prime},z^{\prime})\hat{\rm{\mathbf{y}}}^{\prime}+B_{z}^{\prime}(x^{\prime},y^{\prime},z^{\prime})\hat{\rm{\mathbf{ z}}}^{\prime}.
\end{equation}
Lastly, we transform the $B$-field components and coordinates from the magnetic frame to the rotational frame using a rotation of axes. We rotate the Cartesian frame through an angle $-\alpha$ (the angle between the $\boldsymbol{\Omega}$ and $\boldsymbol{\mu}$ axes), thereby transforming the $B$-field from the magnetic to the rotational frame (indicated by the unprimed coordinates), letting ${\rm {\mathbf{y}}}\parallel{\rm {\mathbf{y}}}^{\prime}$:
\begin{equation}\label{eq:finalB}
\left[\begin{array}{c}
B_x\\
B_y\\
B_z
\end{array}\right]
=
\left[\begin{array}{ccc}
\cos\alpha & 0 & \sin\alpha\\
0 & 1 & 0\\
-\sin\alpha & 0 & \cos\alpha
\end{array}\right]
\left[\begin{array}{c}
B_x^{\prime}\\
B_y^{\prime}\\
B_z^{\prime}
\end{array}\right].
\end{equation}
This yields
\begin{equation}\label{eq:newB2}
\begin{array}{ccrcrcr}
{\rm {\mathbf{B}}}(x^{\prime},y^{\prime},z^{\prime}) & = & B_{x}(x^{\prime},y^{\prime},z^{\prime}){\hat{\rm {\mathbf{x}}}}^{\prime} & + & B_{y}(x^{\prime},y^{\prime},z^{\prime}){\hat{\rm {\mathbf{y}}}}^{\prime} & + & B_{z}(x^{\prime},y^{\prime},z^{\prime}){\hat{\rm {\mathbf{z}}}}^{\prime}.
\end{array}
\end{equation}
Lastly, we transform the the Cartesian coordinates of the position vector from the magnetic to the rotational frame using a similar rotation matrix as Equation~(\ref{eq:finalB}),  and substitute it into Equation~(\ref{eq:newB2}) to obtain
\begin{equation}\label{eq:xyz}
{\mathbf{B}}(x,y,z)=B_{x}(x,y,z)\hat{\mathbf{x}}+B_{y}(x,y,z)\hat{\mathbf{y}}+B_{z}(x,y,z)\hat{\mathbf{z}}.
\end{equation}

\begin{deluxetable}{lccllccrrllc}
\tabletypesize{\tiny}
\tablecaption{\label{Summary} Best-fit parameters for each $B$-field and geometric model combination}
\tablewidth{0pt}
\tablecolumns{12}
\tablehead{\multicolumn{2}{l}{Combinations} & & \multicolumn{4}{c}{Our Best-fit Parameters} & & & \multicolumn{3}{c}{Other Multi-wavelength Fits} \\
\colhead{Model} & \colhead{$\epsilon$} & \colhead{$\chi^2$} & \colhead{$\alpha$} & \colhead{$\zeta$} & \colhead{\textit{A}} & \colhead{$\Delta\phi_{\rm L}$} & \colhead{$\Delta\xi^2_{\rm B}$} & \colhead{$\Delta\xi^2_{\rm all}$}&
\colhead{$\alpha$} & \colhead{$\zeta$} & \colhead{Reference} \\
 & & \colhead{($\times 10^5$)} & \colhead{($^\circ$)} & \colhead{($^\circ$)} & & & & & \colhead{($^\circ$)} & \colhead{($^\circ$)} & }
\startdata
\sidehead{Static dipole \textit{B}-field:}
TPC & ... & 0.819 & $73_{-2}^{+3}$ & $45_{-4}^{+4}$ & 1.3 & 0.55 &   0.00 & 108.75 &  &  &  \\
OG  & ... & 0.891 & $64_{-3}^{+5}$ & $86_{-1}^{+1}$ & 1.3 & 0.05 &   8.44 & 126.75 &  &  &  \\
\sidehead{RVD \textit{B}-field:}
TPC & ... & 3.278 & $54_{-5}^{+5}$ & $67_{-3}^{+5}$ & 0.5 & 0.05 & 723.50 & 723.50 &  &  &  \\
OG  & ... & 0.384 & $78_{-1}^{+1}$ & $69_{-1}^{+2}$ & 1.3 & 0.00 &   0.00 & 0.00 &  &  &  \\
\sidehead{Offset-PC dipole $B$-field for constant $\epsilon_{\nu}$:}
TPC & 0.00 & 0.819 & $73_{-2}^{+3}$ & $45_{-4}^{+4}$ & 1.3 & 0.55 &  0.00 & 108.75 &  &  & \\
    & 0.03 & 0.834 & $73_{-2}^{+2}$ & $43_{-5}^{+4}$ & 1.3 & 0.55 &  1.76 & 112.50 &  &  & \\
    & 0.06 & 0.867 & $73_{-2}^{+2}$ & $42_{-5}^{+5}$ & 1.3 & 0.55 &  5.63 & 120.75 &  &  & \\
    & 0.09 & 0.882 & $73_{-2}^{+1}$ & $41_{-5}^{+3}$ & 1.3 & 0.55 &  7.39 & 124.50 &  &  & \\
    & 0.12 & 1.000 & $74_{-3}^{+1}$ & $42_{-6}^{+3}$ & 1.4 & 0.55 & 21.22 & 154.00 &  &  & \\
    & 0.15 & 0.948 & $73_{-2}^{+1}$ & $39_{-5}^{+3}$ & 1.4 & 0.55 & 15.12 & 141.00 &  &  & \\
    & 0.18 & 0.969 & $73_{-3}^{+2}$ & $37_{-4}^{+4}$ & 1.3 & 0.55 & 17.58 & 146.25 &  &  & \\
\sidehead{Offset-PC dipole $B$-field for variable $\epsilon_{\nu}$:}
SG & 0.00 & 1.587 & $21_{-3}^{+3}$ & $71_{-1}^{+1}$   & 0.5 & 0.85 & 40.52 & 300.75 &  &  & \\
   & 0.03 & 1.627 & $73_{-1}^{+1}$ & $17_{-3}^{+4}$   & 0.7 & 0.55 & 43.96 & 310.75 &  &  & \\
   & 0.06 & 1.525 & $72_{-1}^{+2}$ & $14_{-1}^{+5}$   & 0.5 & 0.60 & 35.18 & 285.25 &  &  & \\
   & 0.09 & 1.452 & $73_{-1}^{+1}$ & $17_{-1}^{+3}$   & 0.6 & 0.55 & 28.90 & 267.00 &  &  & \\
   & 0.12 & 1.437 & $74_{-1}^{+1}$ & $27_{-7}^{+1}$   & 0.8 & 0.55 & 27.61 & 263.25 &  &  & \\
   & 0.15 & 1.116 & $76_{-1}^{+3}$ & $48_{-11}^{+15}$ & 0.7 & 0.55 &  0.00 & 183.00 &  &  & \\
   & 0.18 & 1.119 & $75_{-1}^{+2}$ & $40_{-4}^{+6}$   & 0.5 & 0.55 &  0.26 & 183.75 &  &  & \\
\sidehead{Offset-PC dipole $B$-field for variable $\epsilon_{\nu}$ ($100E_\parallel$):}
SG & 0.00 & 0.581 & $75_{-1}^{+3}$ & $51_{-5}^{+2}$  & 1.1 & 0.55 &  0.00 &  49.27 &  &  & \\
   & 0.03 & 0.634 & $75_{-2}^{+2}$ & $49_{-5}^{+5}$  & 1.1 & 0.55 &  8.73 &  62.48 &  &  & \\
   & 0.06 & 0.698 & $75_{-3}^{+3}$ & $49_{-6}^{+5}$  & 1.1 & 0.55 & 19.39 &  78.61 &  &  & \\
   & 0.09 & 0.774 & $75_{-3}^{+3}$ & $50_{-9}^{+5}$  & 1.1 & 0.55 & 31.90 &  97.54 &  &  & \\
   & 0.12 & 0.789 & $77_{-3}^{+2}$ & $54_{-8}^{+2}$  & 1.1 & 0.55 & 34.42 & 101.36 &  &  & \\
   & 0.15 & 0.845 & $77_{-4}^{+2}$ & $55_{-14}^{+1}$ & 0.9 & 0.55 & 43.62 & 115.28 &  &  & \\
   & 0.18 & 0.834 & $78_{-2}^{+1}$ & $55_{-5}^{+1}$  & 0.8 & 0.55 & 41.80 & 112.51 &  &  & \\
\hline
\multicolumn{2}{l}{RVM}          		  & & & & & & & & 53               & 59.5                   & 1 \\
\multicolumn{2}{l}{X-ray torus}           & & & & & & & &                  & 63.6$^{+0.07}_{-0.05}$ & 2 \\
\multicolumn{2}{l}{RVD and TPC}   		  & & & & & & & & 62--68           & 64                     & 3 \\
\multicolumn{2}{l}{RVD and OG}    		  & & & & & & & & 75               & 64                     & 3 \\
\multicolumn{2}{l}{RVD and Symmetric SG}   & & & & & & & & 44$^{+4}_{-1}$   & 54$^{+1}_{-5}$     & 4 \\
\multicolumn{2}{l}{RVD and Asymmetric SG}  & & & & & & & & 65$^{+1}_{-2}$   & 65.5$^{+2}_{-1}$       & 4 \\
\multicolumn{2}{l}{RVD and OG}             & & & & & & & & 88$^{+2}_{-3}$   & 66.5$^{+1}_{-1}$       & 4 \\
\multicolumn{2}{l}{FF and Symmetric SG}    & & & & & & & & 15$^{+1}_{-1}$   & 68.5$^{+1}_{-1}$       & 4 \\
\multicolumn{2}{l}{FF and Asymmetric SG}   & & & & & & & & 55$^{+10}_{-20}$ & 54.5$^{+4}_{-14}$      & 4 \\
\multicolumn{2}{l}{FF and OG}     		  & & & & & & & & 80$^{+1}_{-1}$   & 53$^{+1}_{-1}$       & 4 \\
\multicolumn{2}{l}{RVD and PC}             & & & & & & & & 3$^{+2}_{-3}$    & 4$^{+2}_{-2}$          & 5 \\
\multicolumn{2}{l}{RVD and SG}             & & & & & & & & 45$^{+2}_{-2}$   & 69$^{+2}_{-2}$         & 5 \\
\multicolumn{2}{l}{RVD and OG}             & & & & & & & & 71$^{+2}_{-2}$   & 83$^{+2}_{-2}$         & 5 \\
\multicolumn{2}{l}{RVD and OPC}            & & & & & & & & 56$^{+2}_{-2}$   & 77$^{+2}_{-2}$         & 5
\enddata
\tablecomments{The table summarizes the best-fit parameters $\alpha$, $\zeta$, $A$, and $\Delta\phi_{\rm L}$, for each model combination, with the errors on $\alpha$ and $\zeta$ determined by using the $3\sigma$ interval connected contours. We chose a minimum error of $1^{\circ}$ if the confidence contour yielded smaller errors. We included the unscaled $\chi^{2}$ to indicate which geometry yields the optimal fit to the Vela data (i.e., the OG model and RVD $B$-field).}
\tablerefs{(1) \citet{Johnston2005}, (2) \citet{Ng2008}, (3) \citet{Watters2009}, (4) \citet{DeCesar2013}, and (5) \citet{Pierbattista2015}.}
\end{deluxetable}


\begin{figure}
    \centering
	\includegraphics[width=12cm]{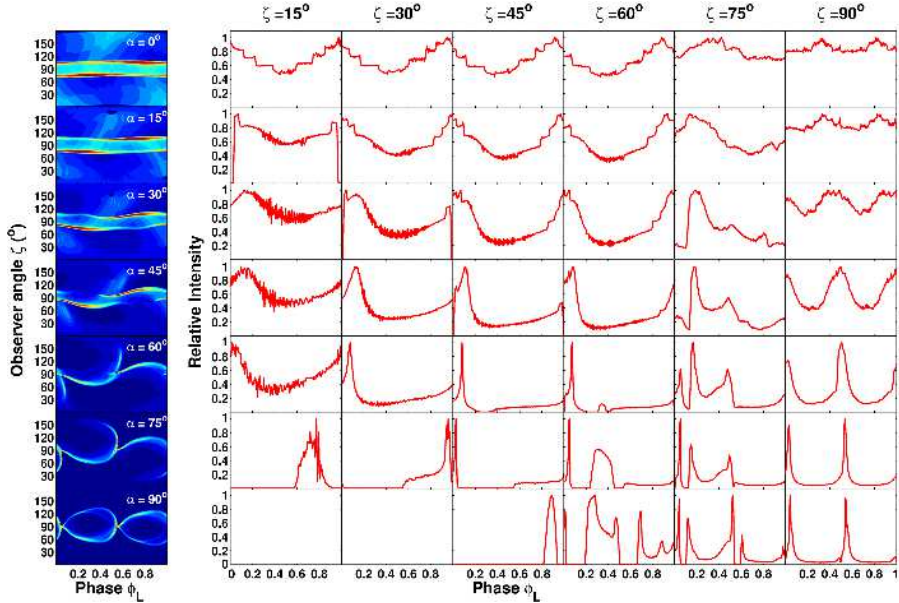}
	\caption{\label{fig:OffsetEps018} The same as in Figure~\ref{fig:OffsetEps000}, but for the TPC model assuming an offset-PC dipole field, for a fixed value of $\epsilon=0.18$ and constant $\epsilon_{\nu}$.}
\end{figure}

\begin{figure}
	\centering	
	\includegraphics[width=12cm]{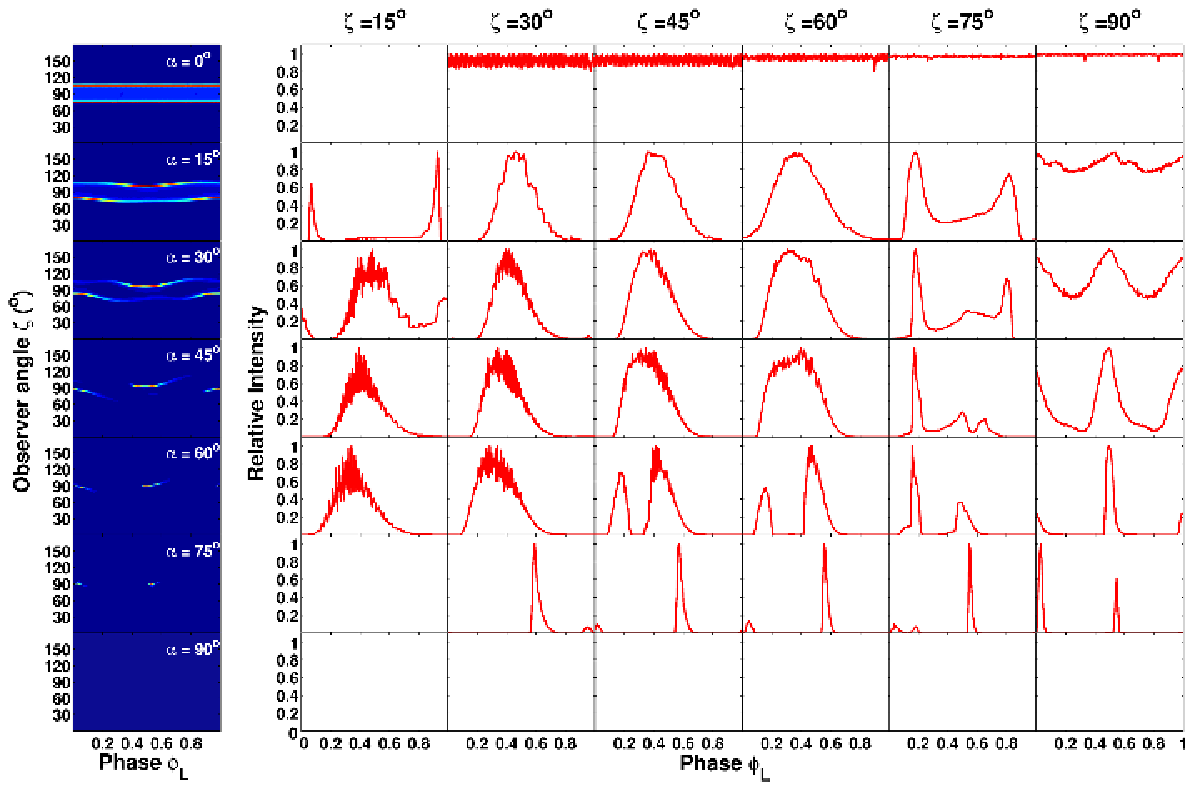}
	\caption{\label{fig:OffsetEps000E} The same as in Figure~\ref{fig:OffsetEps000}, but for the SG model assuming an offset-PC dipole field, for a fixed value of $\epsilon=0.00$ and variable $\epsilon_{\nu}$.}
\end{figure}

\begin{figure}
	\centering
	\includegraphics[width=12cm]{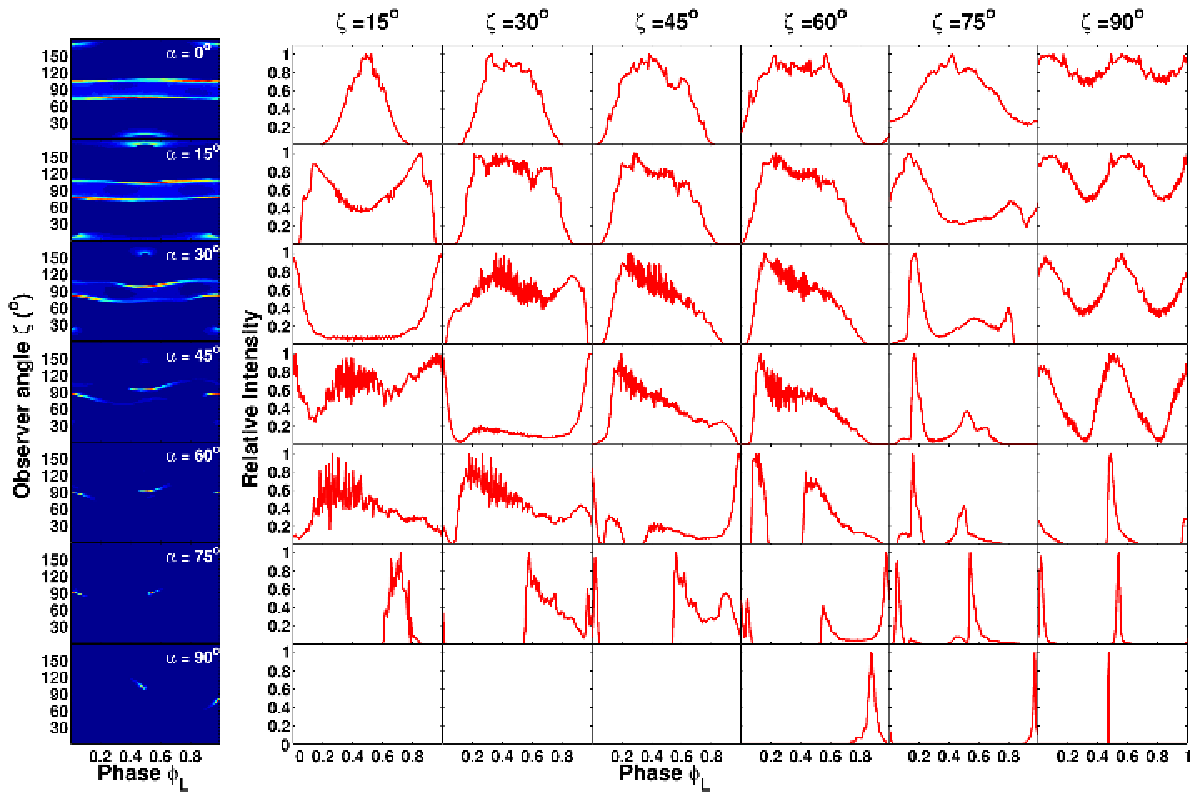}
	\caption{\label{fig:OffsetEps018E} The same as in Figure~\ref{fig:OffsetEps000}, but for the SG model assuming an offset-PC dipole field, for a fixed value of $\epsilon=0.18$ and variable $\epsilon_{\nu}$.}
\end{figure}

\begin{figure}
	\centering
	\includegraphics[width=16cm]{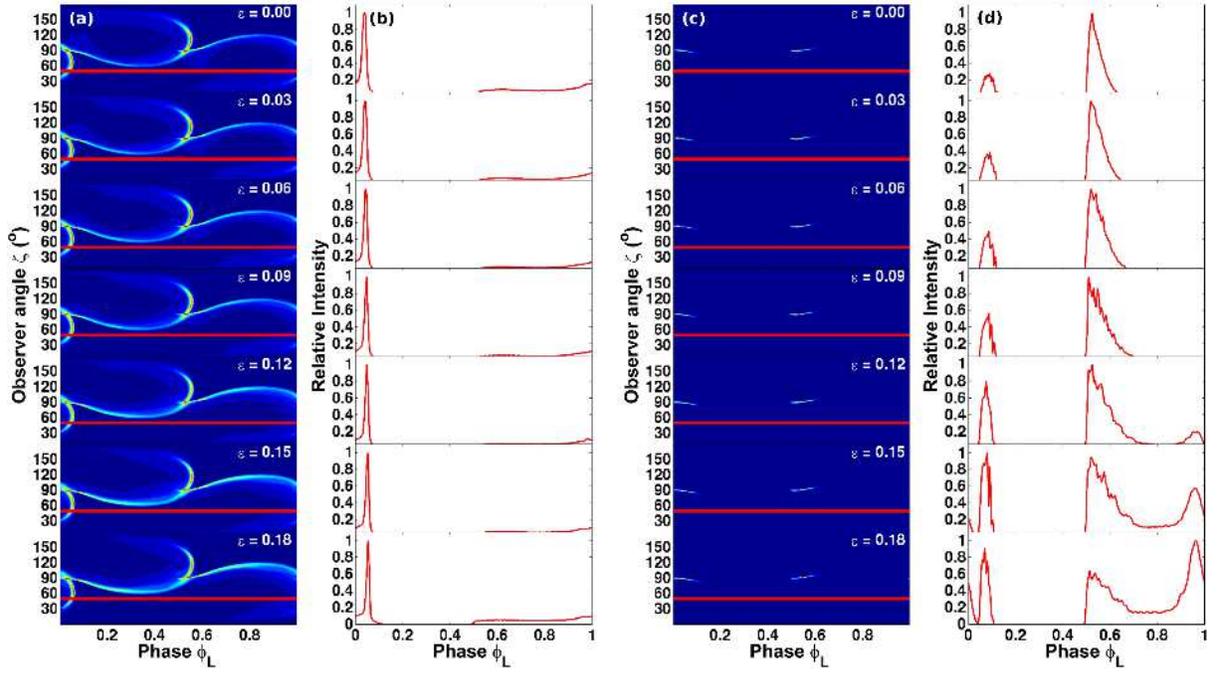}
	\caption{\label{fig:a70z50} Phase plots and light curves for an offset-PC dipole field for $\alpha=70^{\circ}$, $\zeta=50^{\circ}$, and different $\epsilon$ values. Panels (a) and (b) represent the TPC model for constant $\epsilon_{\nu}$, and panels (c) and (d) represent an SG model for variable $\epsilon_{\nu}$.}
\end{figure}

\begin{figure}
	\centering
	\includegraphics[width=12cm]{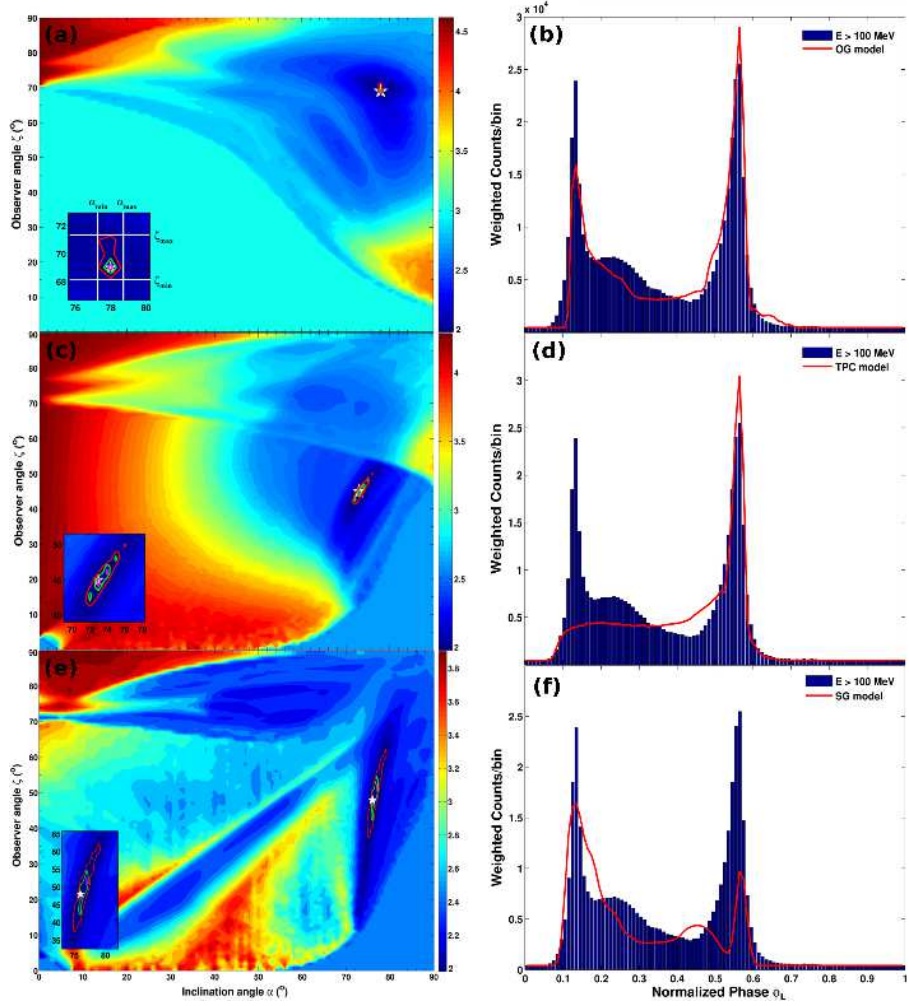}
	\caption{\label{fig:bestcontours} Contour plot for each of the best-fit solutions we obtained for our different $B$-field solutions on an ($\alpha$,$\zeta$) grid. In panel (a) the RVD $B$-field and OG model, panel (c) the offset-PC dipole $B$-field and TPC model (constant $\epsilon_{\rm \nu}$) for $\epsilon=0$, and in panel (e) the offset-PC dipole $B$-field and SG model (variable $\epsilon_{\rm \nu}$) for $\epsilon=0.15$. The color bar of the contour plots represents $\log_{10}\xi^2$, with 1.98 corresponding to the best-fit solution, indicated by the white star. The confidence contour for $1\sigma$ (magenta line), $2\sigma$ (green line), and $3\sigma$ (red line) is also shown with an enlargement in the bottom left corner. The corresponding best-fit light curve for each of the best-fit solutions we obtained for our different $B$-field solutions is also shown. In panel (b) the RVD $B$-field and OG model, panel (d) the offset-PC dipole $B$-field and TPC model (constant $\epsilon_{\rm \nu}$) for an $\epsilon=0$, and in panel (f) the offset-PC dipole $B$-field and SG model (variable $\epsilon_{\rm \nu}$) for an $\epsilon=0.15$. The blue histogram denotes the observed Vela pulsar profile (for energies $E>100$~MeV, \citealp{Abdo2013}) and the red line the model light curve.}
\end{figure}

\begin{figure}
	\centering
	\includegraphics[width=12cm]{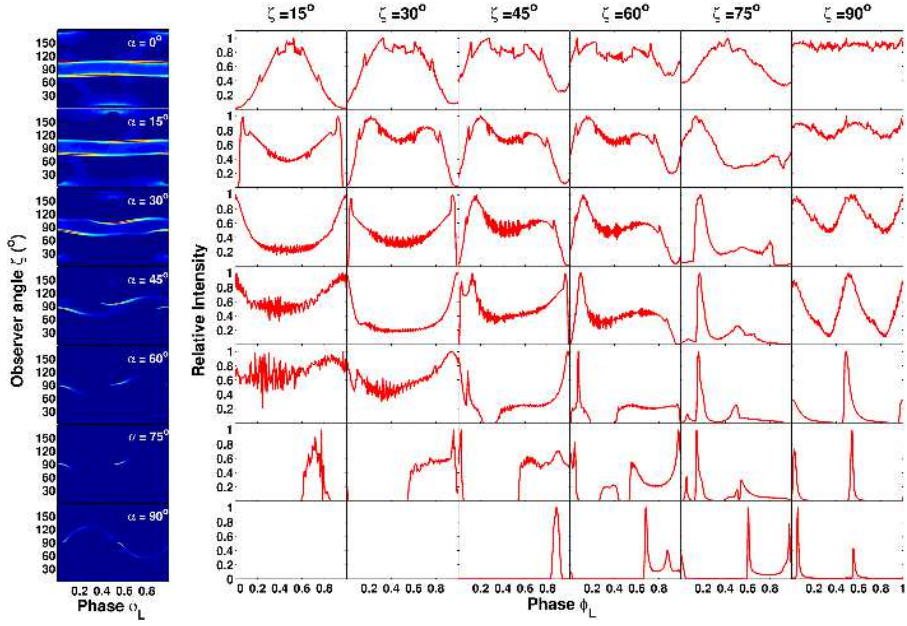}
	\caption{\label{fig:OffsetEps018MeV} The same as in Figure~\ref{fig:OffsetEps018E}, but for a lower $E_{\rm min}$ of 1~MeV.}
\end{figure}

\begin{figure}
	\centering
	\includegraphics[width=12cm]{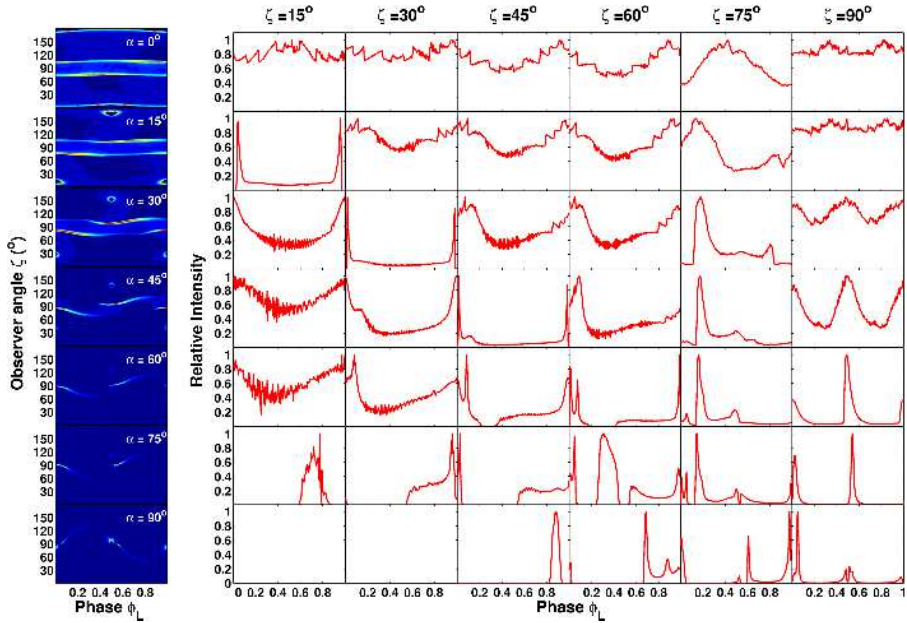}
	\caption{\label{fig:OffsetEps018GeV} The same as in Figure~\ref{fig:OffsetEps018E}, but for the case where we multiplied $E_\parallel$ by a factor of 100, yielding a CR cutoff of $E_{\rm CR}\sim4$~GeV.}
\end{figure}

\begin{figure}
	\centering
	\includegraphics[width=16cm]{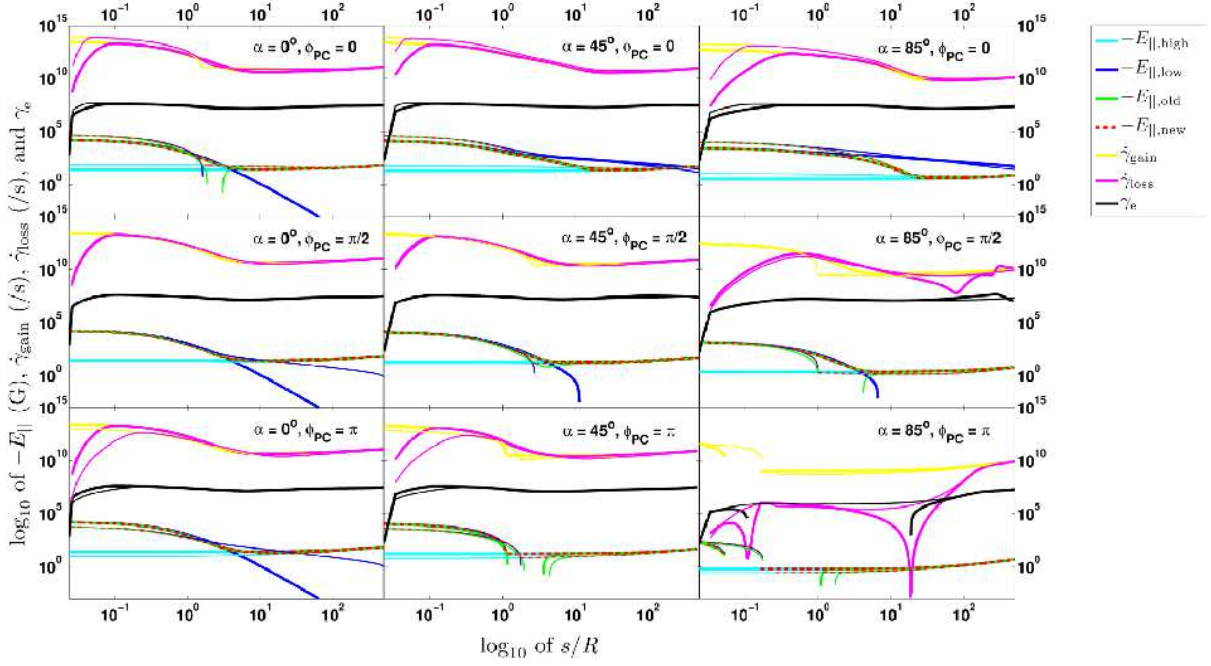}
	\caption{\label{fig:RRLim_100Epar} The same as in Figure~\ref{fig:RRLim}, but for a higher $E$-field, increased by a factor of 100.}
\end{figure}

\begin{figure}
	\includegraphics[width=16cm]{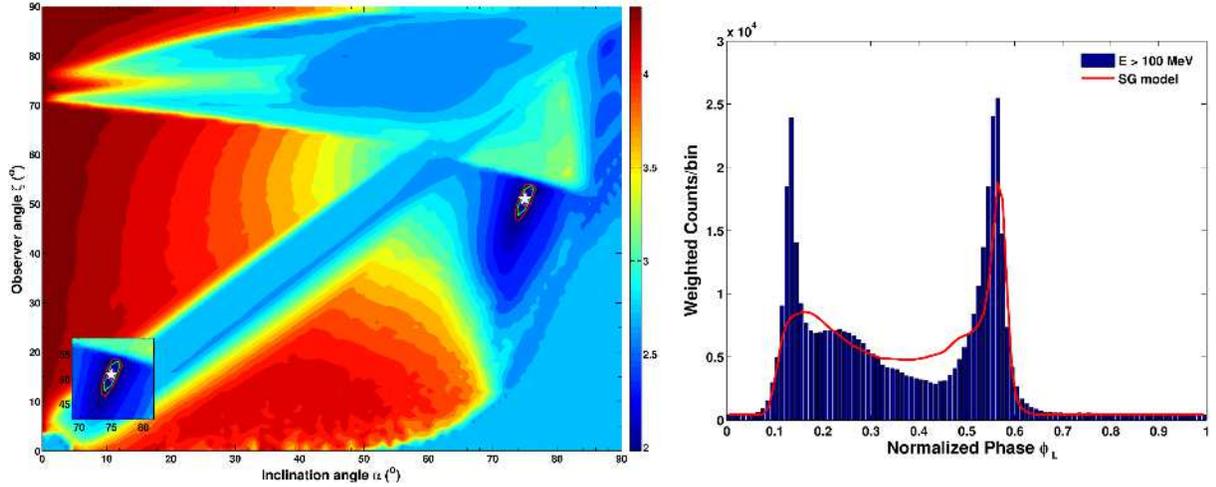}
	\caption{\label{fig:bestcontourLC_Epar} Contour plot (left) and its corresponding best-fit light curve (right) of the best-fit solution we obtained for the offset-PC dipole $B$-field and SG model (with variable $\epsilon_{\rm \nu}$) for $\epsilon=0$, when we multiplied the $E_\parallel$ by a factor of $100$.}
\end{figure}

\begin{figure}
	\centering
	\includegraphics[width=16cm]{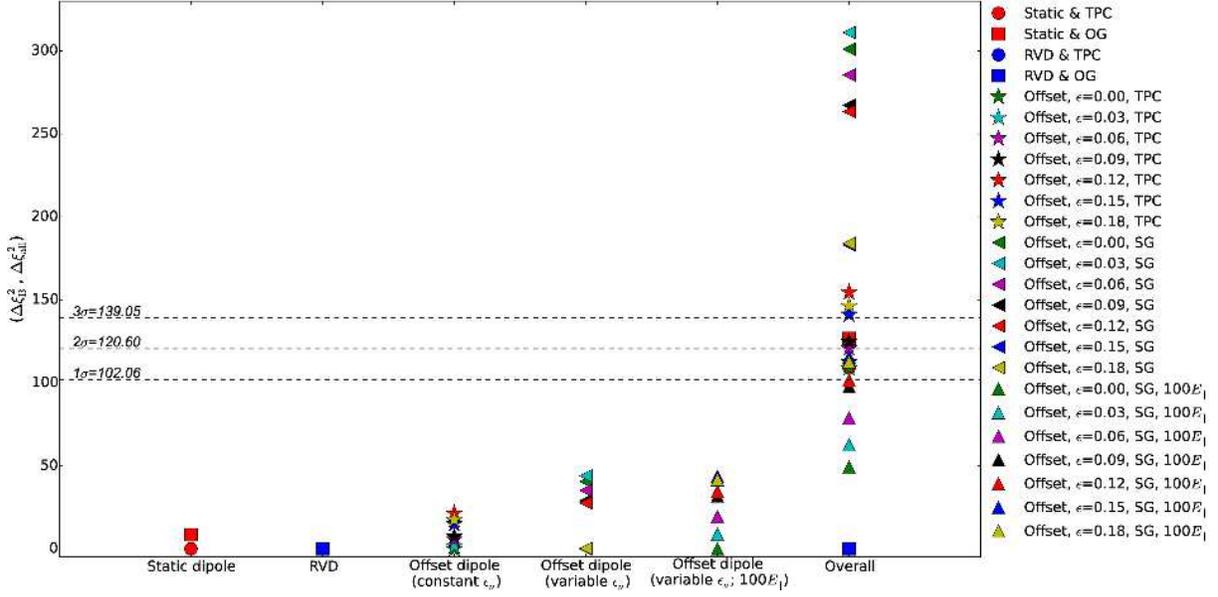}
	\caption{\label{fig:ModelComparisonB} Comparison of the relative goodness of the fit of solutions obtained for each $B$-field and geometric model combination, including the case of $100E_\parallel$, as well as all combinations compared to the overall best fit, i.e., RVD $B$-field and OG model (shown on the $x$-axis). The difference between the optimum and alternative model for each $B$-field is expressed as $\Delta\xi^{2}_{\rm B}$, and for the overall fit as $\Delta\xi^{2}_{\rm all}$ (shown on the $y$-axis). The horizontal dashed lines indicate the $1\sigma$, $2\sigma$, and $3\sigma$ confidence levels. Circles and squares refer to the TPC and OG models for both the static dipole and RVD. The stars refer to the TPC (constant $\epsilon_{{\rm \nu}}$) and the left pointing triangles present the SG (variable $\epsilon_{{\rm \nu}}$) model for the offset-PC dipole field, for the different $\epsilon$ values. The upright triangles refer to our SG model and offset-PC dipole case for a larger $E$-field ($100E_\parallel$). The last column shows our overall fit comparison (see legend for symbols).}
\end{figure}

\begin{figure}
	\centering
	\includegraphics[width=10cm]{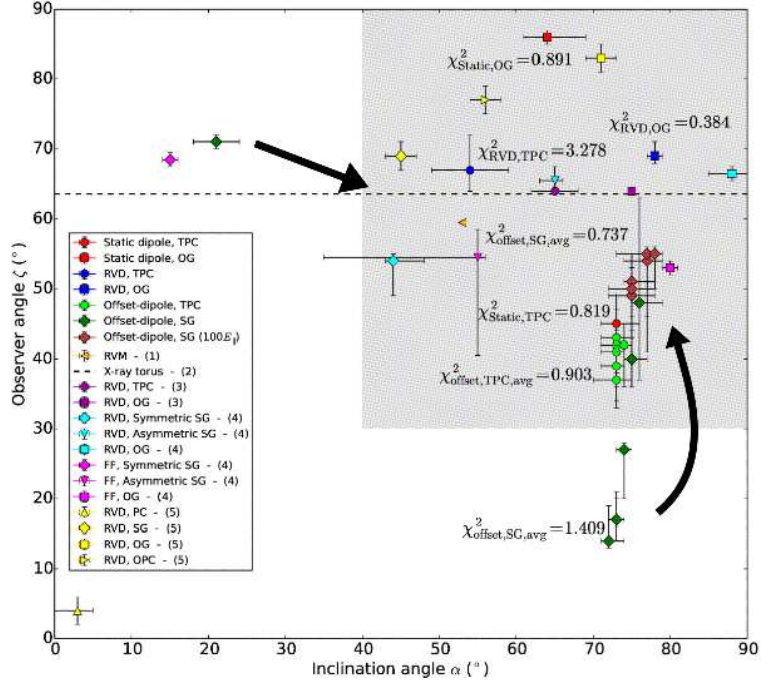}
	\caption{\label{fig:Comparison_alphazeta} Comparison between the best-fit $\alpha$ and $\zeta$, with errors, obtained from this and other studies. Each marker represents a different case as summarized in Table~\ref{Summary}, with the unscaled $\chi^2$~($\times10^5$) value of our fits indicated. For the offset-PC dipole, for both the TPC and SG models we indicate the average $\chi^2$ value over the range of $\epsilon$. We also show our fits for the offset-PC dipole and SG model case with a larger $E_\parallel$. The two black arrows indicate the shift of the best fits to larger $\alpha$ and $\zeta$ if we increase our SG $E$-field by a factor of 100. The shaded region contains all the fits that cluster at larger $\alpha$ and $\zeta$ values.}
\end{figure}

\end{document}